\documentclass[fleqn,usenatbib]{mnras}

\usepackage{mathptmx}
\usepackage[T1]{fontenc}

\DeclareRobustCommand{\VAN}[3]{#2}
\let\VANthebibliography\thebibliography
\def\thebibliography{\DeclareRobustCommand{\VAN}[3]{##3}\VANthebibliography}

\usepackage{graphicx}	% Including figure files
\usepackage{amsmath}	% Advanced maths commands
\usepackage{amssymb}	% Extra maths symbols
\usepackage{xspace}
\usepackage{booktabs}
\usepackage{arydshln}
\usepackage{threeparttable}
\usepackage{hyperref}
\usepackage{orcidlink}

\newcommand{\fref}[1]{Fig.~\ref{#1}}
\newcommand{\tref}[1]{Table~\ref{#1}}
\newcommand{\sref}[1]{Section~\ref{#1}}
\newcommand{\eref}[1]{Eq.~(\ref{#1})}

\newcommand{\tess}{TESS\xspace}
\newcommand{\kepler}{{\it Kepler}\xspace}
\newcommand{\gaia}{{\it Gaia}\xspace}

\newcommand{\Pcapp}{$443$~d\xspace}

\newcommand{\Pb}{$2.699835^{+0.000004}_{-0.000003}$~d\xspace}
\newcommand{\Pc}{$443^{+11}_{-13}$~d\xspace}
\newcommand{\rpb}{$5.24 \pm 0.09$~R$_\oplus$\xspace}
\newcommand{\mpb}{$42 \pm 3$~M$_\oplus$\xspace}
\newcommand{\rhob}{$1.3 \pm 0.5$~g~cm$^{-3}$\xspace}
\newcommand{\mpc}{$84 \pm 7$~M$_\oplus$\xspace}

\newcommand{\eb}{$0.064^{+0.014}_{-0.015}$\xspace}
\newcommand{\ec}{$0.13^{+0.07}_{-0.09}$\xspace}

\newcommand{\teq}{$1266 \pm 27$~K\xspace}

\newcommand{\rprsb}{$0.0476 \pm 0.0005$\xspace}

\newcommand{\rev}[1]{{#1}}

\title[TOI-1288 characterised with HARPS-N and HIRES]{Radial velocity confirmation of a hot super-Neptune discovered by TESS with a warm Saturn-mass companion}

\author[E. Knudstrup et al.]{
E.~Knudstrup\orcidlink{0000-0001-7880-594X},$^{1,2}$\thanks{E-mail: emil@phys.au.dk (EK)}
D.~Gandolfi\orcidlink{0000-0001-8627-9628},$^{3}$
G.~Nowak,$^{4,5}$
C.~M.~Persson\orcidlink{0000-0003-1257-5146},$^{6}$
E.~Furlan\orcidlink{0000-0001-9800-6248},$^{7}$
J.~Livingston,$^{8,9,10}$ \newauthor
E.~Matthews,$^{11}$ 
M.~S.~Lundkvist\orcidlink{0000-0002-8661-2571},$^{1}$
M.~L.~Winther,$^{1}$
J.~L.~Rørsted\orcidlink{0000-0001-9234-430X},$^{1}$
S.~H.~Albrecht\orcidlink{0000-0003-1762-8235},$^{1}$
E.~Goffo\orcidlink{0000-0001-9670-961X},$^{3,12}$\newauthor
I.~Carleo,$^{4}$
H.~J.~Deeg\orcidlink{0000-0003-0047-4241},$^{4,5}$
K.~A.~Collins\orcidlink{0000-0001-6588-9574},$^{13}$
N.~Narita\orcidlink{0000-0001-8511-2981},$^{14,8,4}$
H.~Isaacson,$^{15,7}$
S.~Redfield\orcidlink{0000-0003-3786-3486},$^{16}$
F.~Dai\orcidlink{0000-0002-8958-0683},$^{17,18}$\thanks{NASA Sagan Fellow}\newauthor
T.~Hirano\orcidlink{0000-0003-3618-7535},$^{8,9}$
J.~M.~Akana~Murphy\orcidlink{0000-0001-8898-8284},$^{19}$
C.~Beard\orcidlink{0000-0001-7708-2364},$^{20}$
L.~A.~Buchhave\orcidlink{0000-0003-1605-5666},$^{21}$
S.~Cary\orcidlink{0000-0003-1860-1632},$^{22}$
A.~Chontos\orcidlink{0000-0003-1125-2564},$^{23}$\thanks{Henry Norris Russell Fellow}\newauthor
I.~Crossfield,$^{24}$
W.~D.~Cochran\orcidlink{0000-0001-9662-3496},$^{25}$
D.~Conti\orcidlink{0000-0003-2239-0567},$^{26}$
P.~A.~Dalba\orcidlink{0000-0002-4297-5506},$^{19,27}$\thanks{Heising-Simons 51 Pegasi b Postdoctoral Fellow}
M.~Esposito\orcidlink{0000-0002-6893-4534},$^{12}$
S.~Fajardo-Acosta\orcidlink{0000-0001-9309-0102},$^{7}$\newauthor
S.~Giacalone\orcidlink{0000-0002-8965-3969},$^{15}$
S.~K.~Grunblatt\orcidlink{0000-0003-4976-9980},$^{28}$
P.~Guerra\orcidlink{0000-0002-4308-2339},$^{29}$
A.~P.~Hatzes\orcidlink{0000-0002-3404-8358},$^{12}$
R.~Holcomb\orcidlink{0000-0002-5034-9476},$^{20}$
F.~G.~Horta\orcidlink{0000-0001-9927-7269},\thanks{Citizen Scientist}\newauthor
A.~W.~Howard\orcidlink{0000-0001-8638-0320},$^{18}$
D.~Huber,$^{30}$
J.~M.~Jenkins\orcidlink{0000-0002-4715-9460},$^{31}$
P.~Kab\'{a}th,$^{32}$
S.~Kane\orcidlink{0000-0002-7084-0529},$^{33}$
J.~Korth\orcidlink{0000-0002-0076-6239},$^{6}$
K.~W.~F.~Lam\orcidlink{0000-0002-9910-6088},$^{35}$\newauthor
K.~V.~Lester\orcidlink{0000-0002-9903-9911},$^{31}$
R.~Matson\orcidlink{0000-0001-7233-7508},$^{36}$
K.~K.~McLeod\orcidlink{0000-0001-9504-1486},$^{22}$
J.~Orell-Miquel,$^{4,5}$
F.~Murgas\orcidlink{0000-0001-9087-1245},$^{4,5}$
E.~Palle,$^{4,5}$\newauthor
A.~S.~Polanski\orcidlink{0000-0001-7047-8681},$^{24}$
G.~Ricker\orcidlink{0000-0003-2058-6662},$^{37}$
P.~Robertson\orcidlink{0000-0003-0149-9678},$^{20}$
R.~Rubenzahl\orcidlink{0000-0003-3856-3143},$^{18}$
J~E.~Schlieder,$^{38}$
S.~Seager\orcidlink{0000-0002-6892-6948},$^{39,37,40}$\newauthor
A.~M.~S.~Smith\orcidlink{0000-0002-2386-4341},$^{35}$
P.~Tenenbaum\orcidlink{0000-0002-1949-4720},$^{27,31}$
E.~Turtelboom\orcidlink{0000-0002-1845-2617},$^{15}$
R.~Vanderspek,$^{37}$
L.~Weiss,$^{41}$
and J.~Winn$^{23}$
\\
$^{1}$Stellar Astrophysics Centre, Department of Physics and Astronomy, Aarhus University, Ny Munkegade 120, DK-8000 Aarhus C, Denmark\\
$^{2}$Nordic Optical Telescope, Rambla Jos\'{e} Ana Fern\'{a}ndez P\'{e}rez 7, ES-38711 Bre\~{n}a Baja, Spain\\
$^{3}$Dipartimento di Fisica, Università degli Studi di Torino, via Pietro Giuria 1, I-10125, Torino, Italy \\
$^{4}$Instituto de Astrof\'isica de Canarias (IAC), E-38205 La Laguna, Tenerife, Spain\\
$^{5}$Dept. Astrof\'isica, Universidad de La Laguna (ULL), E-38206 La Laguna, Tenerife, Spain\\
$^{6}$Department of Space, Earth and Environment, Chalmers University of Technology, Onsala Space Observatory, SE-439 92 Onsala, Sweden\\
$^{7}$NASA Exoplanet Science Institute, Caltech/IPAC, Mail Code 100-22, 1200 E. California Blvd., Pasadena, CA 91125, USA\\
$^{8}$Astrobiology Center, 2-21-1 Osawa, Mitaka, Tokyo 181-8588, Japan\\
$^{9}$National Astronomical Observatory of Japan, 2-21-1 Osawa, Mitaka, Tokyo 181-8588, Japan\\
$^{10}$Department of Astronomy, The Graduate University for Advanced Studies (SOKENDAI), 2-21-1 Osawa, Mitaka, Tokyo, Japan\\
$^{11}$Département d’Astronomie, Université de Genève, Chemin Pegasi 51b, 1290 Versoix, Suisse\\
$^{12}$Thüringer Landessternwarte, Tautenburg Sternwarte 5, 07778 Tautenburg, Germany\\
$^{13}$Center for Astrophysics | Harvard \& Smithsonian, 60 Garden Street, Cambridge, MA 02138, USA\\
$^{14}$Komaba Institute for Science, The University of Tokyo, 3-8-1 Komaba, Meguro, Tokyo 153-8902, Japan\\
$^{15}$Department of Astronomy, 501 Campbell Hall, University of California, Berkeley, CA 94720, USA\\
$^{16}$Astronomy Department and Van Vleck Observatory, Wesleyan University, Middletown, CT 06459, USA\\
$^{17}$Division of Geological and Planetary Sciences, 1200 E California Blvd, Pasadena, CA, 91125, USA\\
$^{18}$Department of Astronomy, California Institute of Technology, Pasadena, CA 91125, USA\\
$^{19}$Department of Astronomy and Astrophysics, University of California, Santa Cruz, CA 95064, USA\\
$^{20}$Department of Physics \& Astronomy, University of California Irvine, Irvine, CA 92697, USA\\
$^{21}$DTU Space, National Space Institute, Technical University of Denmark, Elektrovej 328, DK-2800 Kgs. Lyngby, Denmark\\
$^{22}$Department of Astronomy, Wellesley College, Wellesley, MA 02481, USA\\
$^{23}$Department of Astrophysical Sciences, Princeton University, Princeton, NJ 08544, USA\\
$^{24}$Department of Physics and Astronomy, University of Kansas, Lawrence, KS 66045, USA\\
$^{25}$Center for Planetary Systems Habitability and McDonald Observatory, The University of Texas at Austin, Austin TX USA 78712\\
$^{26}$American Association of Variable Star Observers, 185 Alewife Brook Parkway, Suite 410, Cambridge, MA 02138, USA\\
$^{27}$SETI Institute, Carl Sagan Center, 339 Bernardo Ave, Suite 200, Mountain View, CA 94043, USA\\
$^{28}$Department of Physics and Astronomy, Johns Hopkins University, 3400 N Charles St, Baltimore, MD 21218, USA\\
$^{29}$Observatori Astronòmic Albanyà, Camí de Bassegoda S/N, Albanyà 17733, Girona, Spain\\
$^{30}$Institute for Astronomy, University of Hawai‘i, 2680 Woodlawn Drive, Honolulu, HI 96822, USA\\
$^{31}$NASA Ames Research Center, Moffett Field, CA 94035, USA\\
$^{32}$Astronomical Institute of the Czech Academy of Sciences, Fri\v{c}ova 298, 2516, Ond\v{r}ejov\\
$^{33}$Department of Earth and Planetary Sciences, University of California, Riverside, CA 92521, USA\\
$^{34}$Department of Space, Earth and Environment, Chalmers University of Technology, Chalmersplatsen 4, 412 96 Gothenburg, Sweden\\
$^{35}$Institute of Planetary Research, German Aerospace Center (DLR), Rutherfordstrasse 2, D-12489 Berlin, Germany\\
$^{36}$U.S. Naval Observatory, Washington, D.C. 20392, USA\\
$^{37}$Department of Physics and Kavli Institute for Astrophysics and Space Research, Massachusetts Institute of Technology, Cambridge, MA 02139, USA\\
$^{38}$Exoplanets and Stellar Astrophysics Laboratory, NASA Goddard Space Flight Center, 8800 Greenbelt Road, Greenbelt, MD, USA\\
$^{39}$Department of Earth, Atmospheric, and Planetary Sciences, Massachusetts Institute of Technology, Cambridge, MA 02139, USA\\
$^{40}$Department of Aeronautics and Astronautics, Massachusetts Institute of Technology, Cambridge, MA 02139, USA\\
$^{41}$Department of Physics, University of Notre Dame, Notre Dame, IN 46556, USA\\
}

\date{Accepted XXX. Received YYY; in original form ZZZ}

\pubyear{2015}

\begin{document}
\label{firstpage}
\pagerange{\pageref{firstpage}--\pageref{lastpage}}
\maketitle

% Abstract of the paper
\begin{abstract}
We report the discovery and confirmation of the planetary system TOI-1288. This late G dwarf harbours two planets: TOI-1288~b and TOI-1288~c. We combine \tess space-borne and ground-based transit photometry with HARPS-N and HIRES high-precision Doppler measurements, which we use to constrain the masses of both planets in the system and the radius of planet b. TOI-1288~b has a period of \Pb, a radius of \rpb, and a mass of \mpb, making this planet a hot transiting super-Neptune situated right in the Neptunian desert. This desert refers to a paucity of Neptune-sized planets on short period orbits. Our 2.4-year-long Doppler monitoring of TOI-1288 revealed the presence of a Saturn-mass planet on a moderately eccentric orbit (\ec) with a minimum mass of \mpc and a period of \Pc. The 5 sectors worth of \tess data do not cover our expected mid-transit time for TOI-1288~c, and we do not detect a transit for this planet in these sectors.
\end{abstract}

% Select between one and six entries from the list of approved keywords.
% Don't make up new ones.

%%%%%%%%%%%%%%%%%%%%%%%%%%%%%%%%%%%%%%%%%%%%%%%%%%

%%%%%%%%%%%%%%%%% BODY OF PAPER %%%%%%%%%%%%%%%%%%

\section{Introduction}

As the tally of exoplanets has now surpassed 5,000, we can make more informed inferences about planet formation and evolution. A wealth of architectures and different planet types have been discovered, some of which are quite different from the planets found in the Solar System. We first learned about giant planets on short period orbits, the so-called hot Jupiters, which have been found in abundance, owing to their detection bias. The \kepler space mission \citep{Borucki2010} showed us that, while super-Earths appear to be quite common \citep{Howard2010,Mayor2011}, we see a significant dearth of Neptune mass planets on short period orbits, a paucity referred to as the Neptunian ``desert'' \citep{Mazeh2016}.

In addition to this paucity, studies on the planetary initial mass function \citep[e.g.,][]{Mordasini2009} have found a minimum in the mass range where super-Neptunes reside, namely from around 30~M$_\oplus$ to 70~M$_\oplus$. This valley has been interpreted as the division between planets dominated by solids and gas giants that have undergone runaway gas accretion \citep{Ida2004}. Finding and characterising planets in this mass range could therefore help shed light on why some proto-planets undergo runaway accretion while others do not.

Most of the super-Neptunes were detected by \kepler around relatively faint stars, meaning that precise mass determinations only exist for a few of these \citep[e.g., Kepler-101b,][]{Bonomo2014}. The Transiting Exoplanet Survey Satellite \citep[TESS;][]{art:ricker2015} along with ground-based efforts have now detected more of these super-Neptunes in brighter systems for which precise radial velocities (RVs) are more viable, enabling both radius and mass determinations. Therefore, we can also determine the bulk density and make inferences about the composition. A way to gain more insight into the composition and potential migration is through atmospheric studies, which have also been used as a means to rule out certain mechanisms. For instance, as in \citet{Vissapragada2022} in which photoevaporation is ruled out as the mechanism responsible for shaping the upper edge of the Neptunian desert.

Here we report on the discovery and characterisation of the TOI-1288 planetary system. In this system we have discovered a hot super-Neptune, TOI-1288~b, with an outer Saturn mass companion, TOI-1288~c. These planets are hosted by a late G dwarf.

The paper is structured as follows. In \sref{sec:obs} we describe our observations, which include ground-based photometry as well as that from \tess. We have also acquired speckle and adaptive optics (AO) imaging to search for blended companions. In addition we have carried out extensive spectroscopic follow-up to confirm and characterise this planetary system. In \sref{sec:analysis} we present our analysis of the data in which we model the photometry and spectroscopy jointly. The results are presented in \sref{sec:results} and  we discuss them in \sref{sec:disc}. Finally, we give our conclusions in \sref{sec:conc}.

\begin{table}
    \centering
    \caption{ {\bf System parameters.} Catalog IDs, coordinates, and magnitudes for the TOI-1288 system.}
    \begin{threeparttable}
    \begin{tabular}{c c c}
        \toprule 
        Parameter & Value & Name \\
        \midrule
         TIC\tnote{a} & 365733349 & \\
         {\it Gaia} DR3\tnote{b} & 2245652826430109184 & \\
         TYC\tnote{c} & 4255-1629-1 & \\
         \midrule
         $\alpha$ (J2000)\tnote{b} & 20:52:40.09 & Right ascension (R.A.) \\
         $\delta$ (J2000)\tnote{b} & +65:36:31.59 & Declination (Dec.) \\
         $\mu_\alpha$ (mas~yr$^{-1}$)\tnote{b} & $43.496 \pm 0.017$ & Proper motion R.A. \\
         $\mu_\delta$ (mas~yr$^{-1}$)\tnote{b} & $-68.775 \pm 0.017$ & Proper motion Dec. \\
         $\varpi$ (mas)\tnote{b} & $8.720 \pm 0.013$ & Parallax \\
         RV (km~s$^{-1}$)\tnote{b} & $-68.1 \pm 0.6$ & Radial velocity \\
         \midrule
         $G$\tnote{b} & $10.4507 \pm 0.0018$ & {\it Gaia} $G$ magnitude\\
         $B_\mathrm{P}$\tnote{b} & $10.855 \pm 0.006$ & {\it Gaia} $B_\mathrm{P}$ magnitude\\
         $R_\mathrm{P}$\tnote{b} & $9.873 \pm 0.003$ & {\it Gaia} $R_\mathrm{P}$ magnitude\\
         $V$\tnote{c} & $10.44\pm 0.04$ & Tycho $V$ magnitude\\
         $B$\tnote{c} & $11.38 \pm 0.07$ & Tycho $B$ magnitude\\
         $J$\tnote{d} & $9.19 \pm 0.02$ & 2MASS $J$ magnitude\\
         $H$\tnote{d} & $8.84 \pm 0.03$ & 2MASS $H$ magnitude\\
         $K$\tnote{d} & $8.78 \pm 0.02$ & 2MASS $K$ magnitude\\
         \bottomrule
    \end{tabular}
\begin{tablenotes}
    \item[a] \url{https://exofop.ipac.caltech.edu/tess/}.
    \item[b] \citet{GaiaDR3}.
    \item[c] \citet{art:hog2000}.
    \item[d] \citet{Cutri2003}.
\end{tablenotes}
\end{threeparttable}
    \label{tab:star}
\end{table}

\section{Observations}
\label{sec:obs}

The TOI-1288 system has been observed with different space- and ground-based facilities, including both photometric and spectroscopic observations, as well as high-resolution imaging. System parameters for TOI-1288 are summarised in \tref{tab:star}.

\subsection{Photometry}
\label{sec:tess}
\tess observed TOI-1288 during Sectors 15, 16, 17, 18, and 24 (August 15 to November 27, 2019, and April 16 to May 13, 2020). This candidate was identified by the Science Processing Operation Center \citep[SPOC;][]{Jenkins2016} team at the NASA Ames Research Center, who searched the light curves, which are extracted through simple aperture photometry \citep[SAP;][]{Twicken2010,Morris2020} and processed using the Presearch Data Conditioning \citep[PDC;][]{Smith2012,Stumpe2012,Stumpe2014} algorithm. The SPOC team searches the PDC-SAP light curves for transit-like signals with an adaptive, noise-compensating matched filter \citep{Jenkins2002,Jenkins2010} using a pipeline that iteratively performs multiple transiting planet searches and stops when it fails to find subsequent transit-like signatures above the detection threshold of a signal-to-noise ratio (SNR) of 7.1. The results were published in the Data Validation Report \citep[DVR;][]{Twicken2018,Li2019}, and as the light curve shows a $\sim$0.25\% dip occurring every 2.7~d with an SNR of around 62, it was identified as a TESS Object of Interest \citep[TOI;][]{art:guerrero2021} and given the ID TOI-1288. The results of the difference image centroiding test were also presented in the DVR, which located the source of the transit signal to within $1.3\pm2.6^{\prime \prime}$ in the Sector 14-26 multi-sector transit search.

\begin{figure}
    \centering
    \includegraphics[width=\columnwidth]{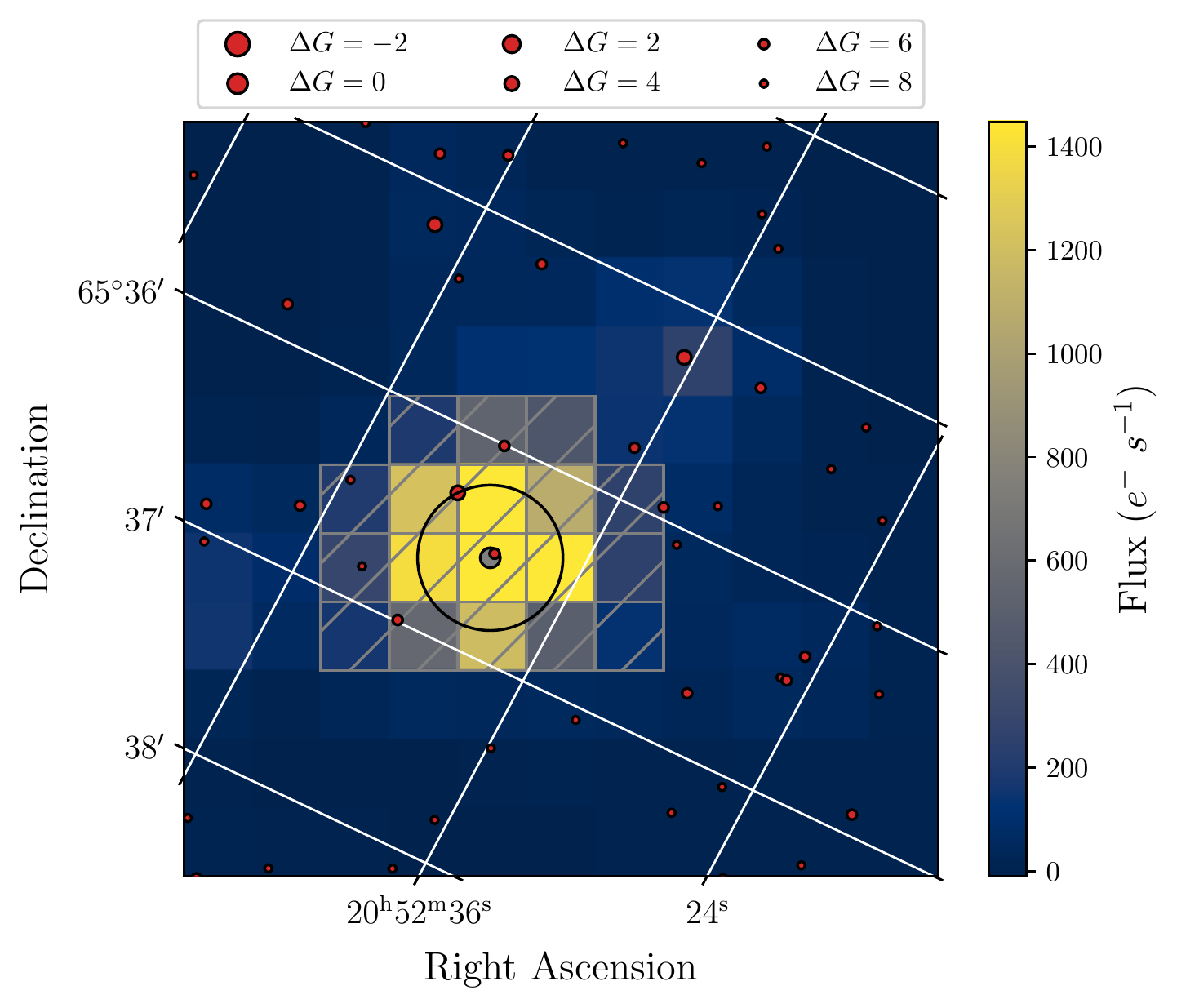}
    \caption{{\bf \tess image of TOI-1288.} Cutout of a TESS image of TOI-1288 from Sector 15. The red dots denote \gaia sources with their sizes scaled to the difference in $G$ magnitude to TOI-1288. The grey dot denotes the position of TOI-1288. The hatched area shows the aperture mask we used to create the light curves, and the black circle illustrates the separation to the brightest nearby star at $\sim$23~arcsec.}
    \label{fig:tpf}
\end{figure}

An independent search for transit signals was performed using the Détection Spécialisée de Transits \citep[DST;][]{Cabrera2012} pipeline on the PDCSAP light curves. A transit signal with orbital period of $2.70 \pm 0.02$ days and a transit depth of $\sim$0.25\% was detected, consistent with the signal detected by the SPOC pipeline.

\begin{figure*}
    \centering
    \includegraphics[width=\textwidth]{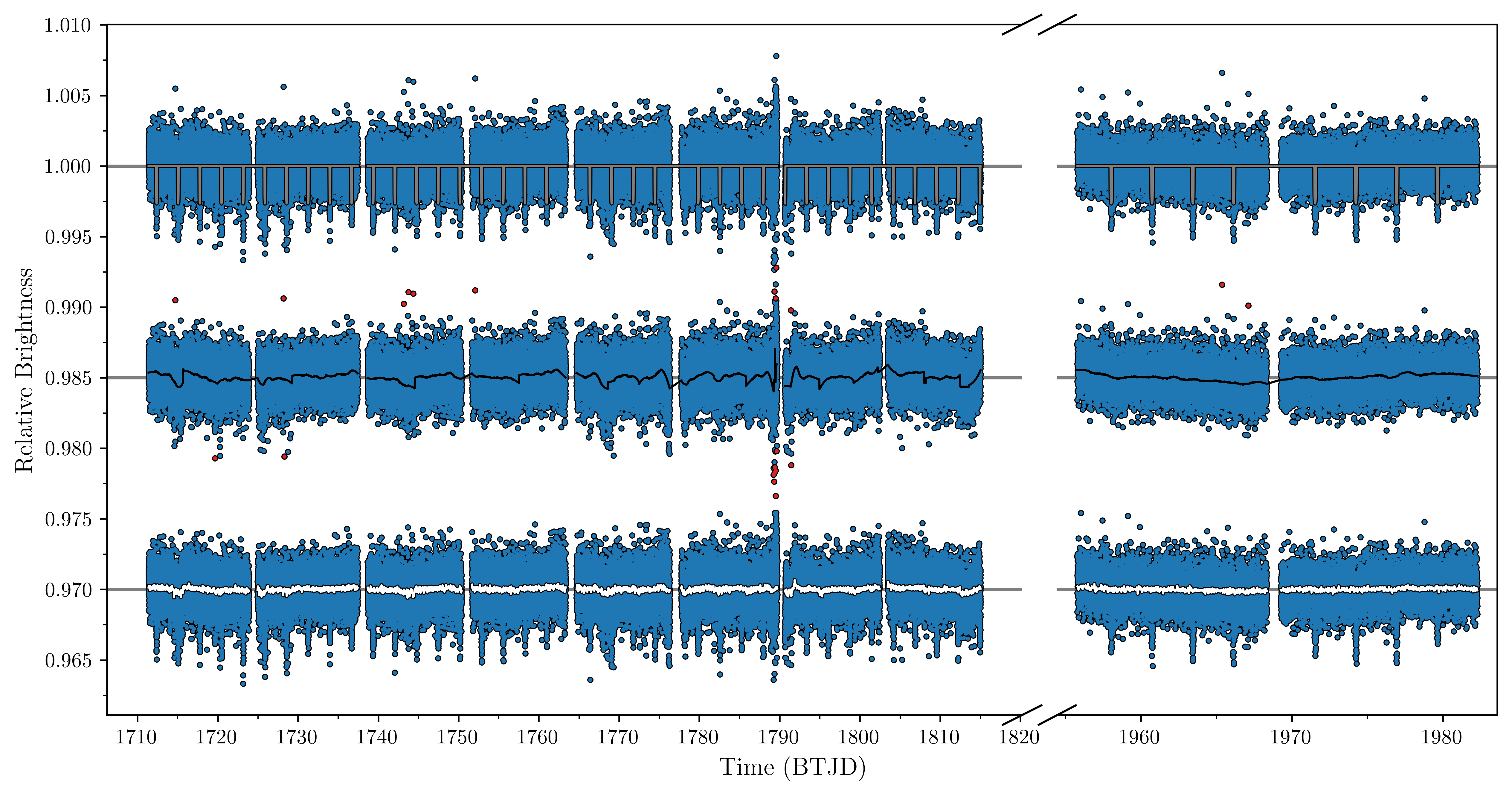}
    \caption{{\bf \tess photometry.} The light curve at the top shows the background corrected light curve. The grey line is a transit model created from parameters stemming from an initial fit. The transit model has been used to temporarily remove the transit in the light curve shown in the middle. Here the grey line shows a Savitsky-Golay filter \citep[as implemented in][]{misc:lightkurve} used to filter and detrend the data for outlier rejection. The red points are outliers removed through a 5$\sigma$ sigma clipping. The TESS data with outliers removed and the transits re-injected is shown in the light curve at the bottom. The white line is the GP we use to detrend the data (see \sref{sec:analysis}).
    }
    \label{fig:tess}
\end{figure*}

\fref{fig:tpf} displays the \tess image in the immediate vicinity of TOI-1288 \rev{with nearby \gaia DR3 sources \citep{GaiaDR3}}. All the \tess photometry from Sectors 15-18 and Sector 24 is displayed in \fref{fig:tess}, where we show the background corrected light curve at the top. This was done using the \texttt{RegressionCorrector} implemented in \texttt{lightkurve} \citep{misc:lightkurve}. Overplotted in grey is a model light curve created using \texttt{batman} \citep{art:kreidberg2015} with transit parameters stemming from an initial fit. We used this to remove the transit signal before removing outliers from the light curve. In the middle light curve the transits are removed, and we have applied a Savitzky-Golay filter \citep{art:savitzky1964} to temporarily filter the light curve. We then removed outliers through sigma clipping at 5$\sigma$, these outliers are highlighted in red. Finally, in the bottom light curve we have re-injected the transits to the unfiltered light curve as we want to account for any trend while fitting, as described in \sref{sec:analysis}.

\subsubsection{Light Curve Follow-up}\label{sec:lcfup}

We acquired ground-based time-series follow-up photometry of TOI-1288 as part of the \tess Follow-up Observing Program \citep[TFOP;][]{Collins2019}\footnote{\url{https://tess.mit.edu/followup}.} \rev{using various facilities as listed in \tref{tab:transitfollowup} from October 2019 to September 2021}. This is done in an attempt to (1) rule out or identify nearby eclipsing binaries (NEBs) as potential sources of the detection in the \tess data, (2) detect the transit-like events on target to confirm the depth and thus the \tess photometric deblending factor, (3) refine the \tess ephemeris, and (4) place constraints on transit depth differences across optical filter bands. We used the {\tt TESS Transit Finder}, which is a customized version of the {\tt Tapir} software package \citep{Jensen2013}, to schedule our transit observations. The images were calibrated and the photometric data were extracted using the {\tt AstroImageJ} ({\tt AIJ}) software package \citep{Collins2017}, except the Las Cumbres Observatory Global Telescope \citep[LCOGT;][]{Brown2013} images, which were calibrated by the standard LCOGT {\tt BANZAI} pipeline \citep{McCully2018}, and the Multicolor Simultaneous Camera for studying Atmospheres of Transiting exoplanets \citep[MuSCAT;][]{Narita2015} data, which were extracted using the custom pipeline described in \citep{Fukui2011}.

The individual observations are detailed in \tref{tab:transitfollowup} and the light curves are shown in \fref{fig:lcb}. All photometric apertures exclude flux from all known \gaia DR3 stars near TOI-1288, except the \tess-band 16.4 magnitude neighbor 1.5\arcsec southwest, which is nominally too faint to be capable of causing the detection in the \tess photometric aperture (individual follow-up photometric apertures are listed in \tref{tab:transitfollowup}). Transit events consistent with the \tess TOI-1288~b transit signal were detected in each light curve and are included in the joint model described in \sref{sec:analysis}.

\subsection{Speckle/AO imaging}
\label{sec:speck}

Nearby sources that are blended in the aperture mask used for the photometry can contaminate the light curve and alter the measured radius, it is thus important to vet for close visual companions. Furthermore, a close companion could be the cause of a false positive if the companion is itself an eclipsing binary \citep{Ciardi2015}.  We therefore collected both adaptive optics and speckle imaging. The observations are described below and summarised in \tref{tab:speck}.

\subsubsection{WIYN/NESSI}

\begin{figure}
    \centering
    \includegraphics[width=\columnwidth]{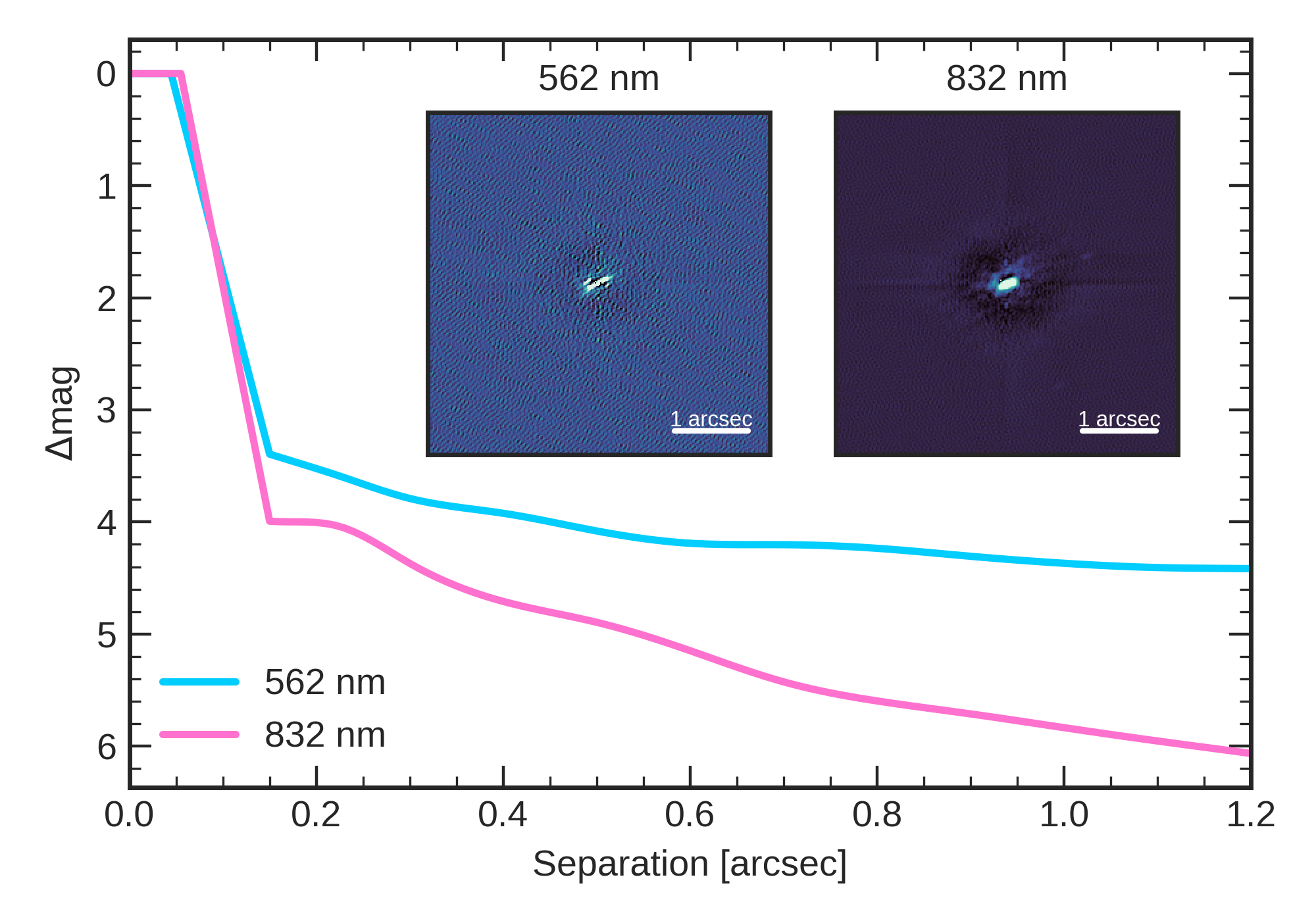}
    \includegraphics[width=\columnwidth]{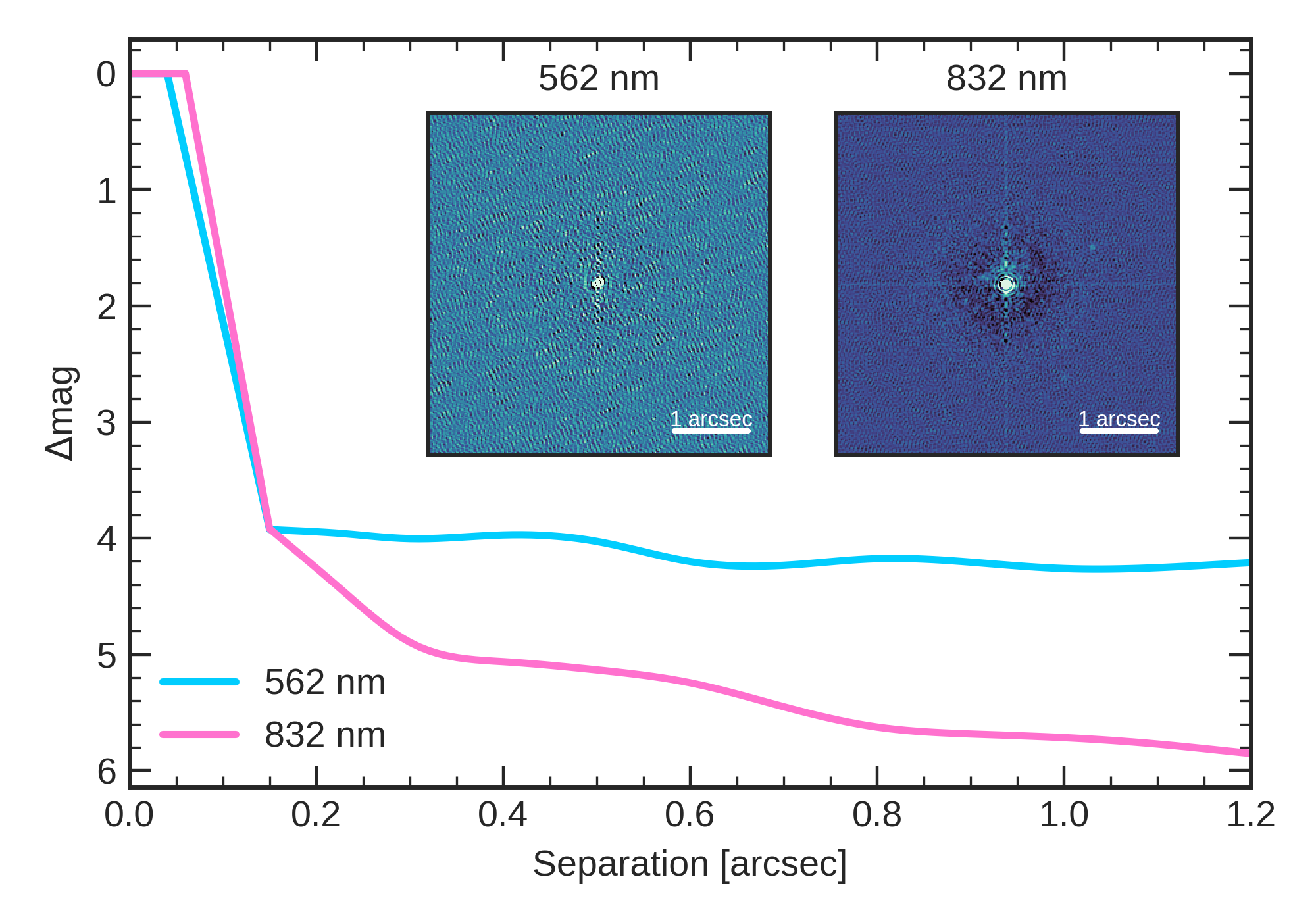}
    \caption{{\bf WIYN/NESSI contrast curves from 2019 (top) and 2021 (bottom).} Two filter speckle imaging contrast curves for TOI-1288 from NESSI. The insets show the reconstructed 562~nm and 832~nm images with 1~arcsec scale bars.}
    \label{fig:wiyn}
\end{figure}

On the nights of 2019 November 17 and 2021 October 29, TOI-1288 was observed with the NESSI speckle imager \citep{Scott2019}, mounted on the 3.5\,m WIYN telescope at Kitt Peak, AZ, USA. NESSI simultaneously acquires data in two bands centered at 562\,nm and 832\,nm using high speed electron-multiplying CCDs (EMCCDs). We collected and reduced the data following the procedures described in \citet{Howell2011}. The resulting reconstructed image achieved a contrast of $\Delta\mathrm{mag}\,\approx\,5.75$ at a separation of 1\arcsec in the 832\,nm band (see \fref{fig:wiyn}). 

On both nights we detected a companion at a separation of $\sim$1.2\arcsec, however, only in the 832~nm filter. Additionally, on the night of 2021 October 29 the pipeline detected a companion at a separation of 0.065\arcsec (position angle of $313^\circ$ and $\Delta\mathrm{mag} =2.57$). However, this companion is close to the detection limit \citep{Scott2019}, and the fit that produced it relied on image elongation (as opposed to being fully separated from the primary), which is possible to get from a mismatch between the science target and the (single) comparison star. Furthermore, it was not detected in the 2019 November 17 data (despite being of higher quality), nor was it detected in any of the other speckle or AO images (see below). We therefore conclude that the inner companion is a spurious detection most likely caused by a data artifact.

\subsubsection{Gemini/'Alopeke}

TOI-1288 was observed with the 'Alopeke speckle instrument on the Gemini North telescope, HI, USA, \citep{Scott2021} on 2020 June 9, 2021 June 24, 2021 October 22, and 2022 May 14 (all dates in UT). Observations were obtained simultaneously in two narrow-band filters centered at 562~nm (width=54~nm) and at 832~nm (width=40~nm). Between 6 and 7 sets of 1000$\times$0.06~s exposures were collected and then reduced with the standard reduction pipeline using Fourier analysis \citep[see, e.g.,][for an overview]{Howell2011}. The reduced data products include reconstructed images and 5$\sigma$ contrast curves. TOI-1288 was very faint in most data sets and even not detected in one of them at 562~nm. At 832~nm, in addition to the primary star, a faint \rev{($\Delta M \sim 5.9 $)} companion was detected at a projected separation of $\sim$1.2$^{\prime\prime}$-1.3$^{\prime\prime}$ in the data from 2020 June 9, 2020 June 24, 2021 October 22, and 2022 May 14. An even fainter \rev{($\Delta M \sim 7$)} companion was detected at a separation of $\sim$1.4$^{\prime\prime}$-1.5$^{\prime\prime}$ in the data from 2020 June 24, 2021 October 22, and 2022 May 14.

\subsubsection{Gemini/NIRI}

We collected adaptive optics images of TOI-1288 with the Gemini Near-Infrared Imager \citep[NIRI;][]{Hodapp2003} on 2019 November 8. We collected 9 science frames, each with an exposure time of 6.8~s, and dithered the telescope by $\sim$2\arcsec\ between each frame, thereby allowing for the science frames themselves to serve as sky background frames. The target was observed in the Br-$\gamma$ filter centered at 2.166~$\mu$m. Data processing consisted of bad pixel removal, flat fielding, and subtraction of the sky background. We then aligned the frames based on the position of the primary star, and coadded the images.

The total field of view is around 26$^{\prime\prime}$ square, with optimum sensitivity in the central $\sim$22$^{\prime\prime}$ square. We again identified two visual candidates in the field of view. The brighter companion is at a separation of 1.152$^{\prime\prime}$, a position angle of 289.3~degrees \rev{counter-clockwise} of north, and is 4.77$\pm$0.03~mag fainter than the host in the Br-$\gamma$ band; the fainter companion is at a separation of 1.579$^{\prime\prime}$, a position angle of 207.7~degrees \rev{counter-clockwise} of north and is 5.88$\pm$0.04~mag fainter than the host.

\begin{figure}
    \centering
    \includegraphics[width=\columnwidth]{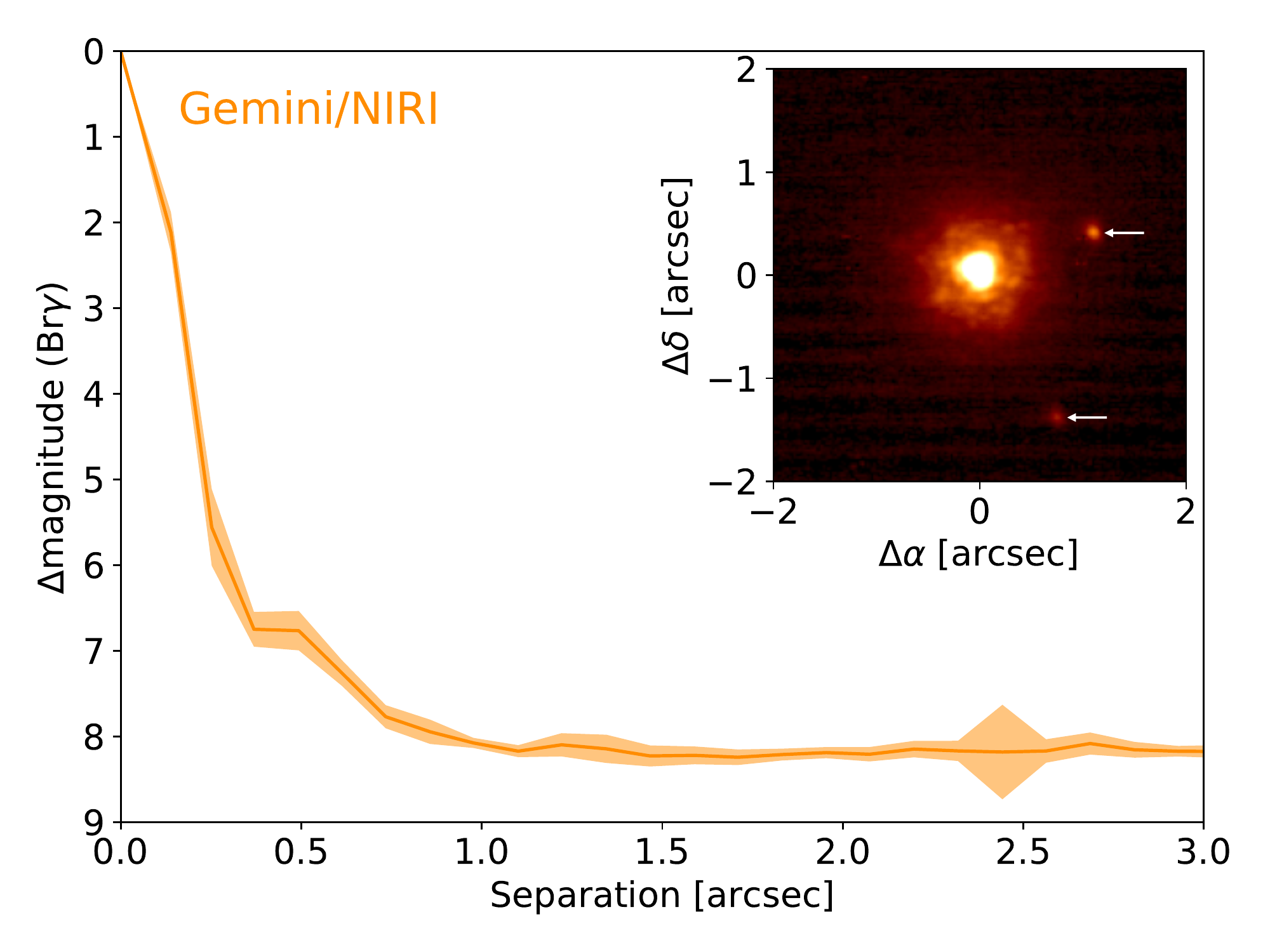}
    \caption{{\bf Gemini/NIRI contrast curve.} AO imaging contrast curve for TOI-1288. The inset shows the reconstructed Br-$\gamma$ image with the two detected companions highlighted.}
    \label{fig:gem_niri}
\end{figure}

We measured the sensitivity of our observations as a function of radius by injecting fake companions and scaling their brightness such that they could be detected at 5$\sigma$. The contrast sensitivity is 5.56~mag fainter than the host at a separation of 250~mas, and 8.1~mag fainter than the host in the background limited regime, beyond $\sim$1$^{\prime\prime}$ from the target. The contrast sensitivity as a function of radius and a high resolution image of the star are shown in \fref{fig:gem_niri}; we show the curve for the inner 3$^{\prime\prime}$ only, but note that the data are sensitive to candidates within 13$^{\prime\prime}$ in all directions. From our speckle and AO imaging we have thus identified two nearby companions.

\begin{table*}
    \centering
    \caption{{\bf Companions detected in Speckle, AO, and \textbf{\textit{Gaia}}.} $\rho$ and $\Delta$ are the separation and the difference in magnitude from the central target (host), respectively. $\theta$ is the position angle from the brighter of the targets to the fainter component, measured from North through East. Star 1 is the brighter companion and star 2 the fainter one. \rev{The uncertainties for $\rho$ and $\theta$ for the 'Alopeke data are estimated to be around 5~mas and 1~deg, respectively, while the uncertainties for $\Delta$mag come out to around 0.5~mag for the closer companion and 1~mag for the fainter one}.
    }
    \begin{threeparttable} 
    \begin{tabular}{c c c c c c c c c}
    \toprule
         Date & Star & $\rho$ & $\Delta$ & $\theta$ & Type & Filter & Instrument & Telescope \\
         (UT) & & (arcsec) & (mag) & (deg) & & & & \\
    \midrule
        2019-11-08 & 1 & 1.152 & $4.77\pm0.03$ & 289.3 & AO & Br-$\gamma$ & NIRI & Gemini \\
        2019-11-08 & 2 & 1.579 & $5.88\pm0.04$ & 207.7 & AO & Br-$\gamma$ & NIRI & Gemini \\
        
        2019-11-17\tnote{a} & 1 & 1.123 & 5.90 & 289.5 & Speckle & 832~nm & NESSI & WIYN \\

        2020-06-09\tnote{a} & 1 & 1.172 & 5.94 & 289.5 & Speckle & 832~nm & 'Alopeke & Gemini \\

        2021-06-24\tnote{a} & 1 & 1.256 & 6.4 & 292.0 & Speckle & 832~nm & 'Alopeke & Gemini \\
        2021-06-24\tnote{a} & 2 & 1.516 & 6.8 & 211.7 & Speckle & 832~nm & 'Alopeke & Gemini \\
        
        2021-10-22\tnote{a} & 1 &  1.233 & 5.92 & 293.4 & Speckle & 832~nm & 'Alopeke & Gemini \\
        2021-10-22\tnote{a} & 2 &  1.468 & 7.34 & 212.3 & Speckle & 832~nm & 'Alopeke & Gemini \\

        2021-10-29\tnote{a} &  1 & 1.282  & 5.48  & 293.6  & Speckle & 832~nm & NESSI & WIYN \\
         
        2022-05-14\tnote{a} & 1 & 1.306 & 5.83 & 293.5 & Speckle & 832~nm & 'Alopeke & Gemini \\
        2022-05-14\tnote{a} & 2 & 1.443 & 7.90 & 214.3 & Speckle & 832~nm & 'Alopeke & Gemini \\

        Epoch=2016.0 & 2 & 1.74 & 6.41 & 198 & Photometry & $G$ & - & \gaia \\
    \bottomrule
    \end{tabular}
    \begin{tablenotes}
        \item[a] Observations were also carried out in the 562~nm filter, but the companions were not detected in this filter.
        
    \end{tablenotes}
    \end{threeparttable}
    \label{tab:speck}
\end{table*}

\subsubsection{\gaia}

As is also apparent from \fref{fig:tpf}, one of the two aforementioned companions is also detected by \gaia DR3. The position of this \gaia companion is in good agreement with it being the fainter of the two companions seen in the Gemini 'Alopeke and AO observations. This is most likely also the companion seen in the light curve follow-up in \sref{sec:lcfup}. The \gaia detection is summarised in \tref{tab:speck} along with the speckle and AO observations. 

\begin{figure}
    \centering
    \includegraphics[width=\columnwidth]{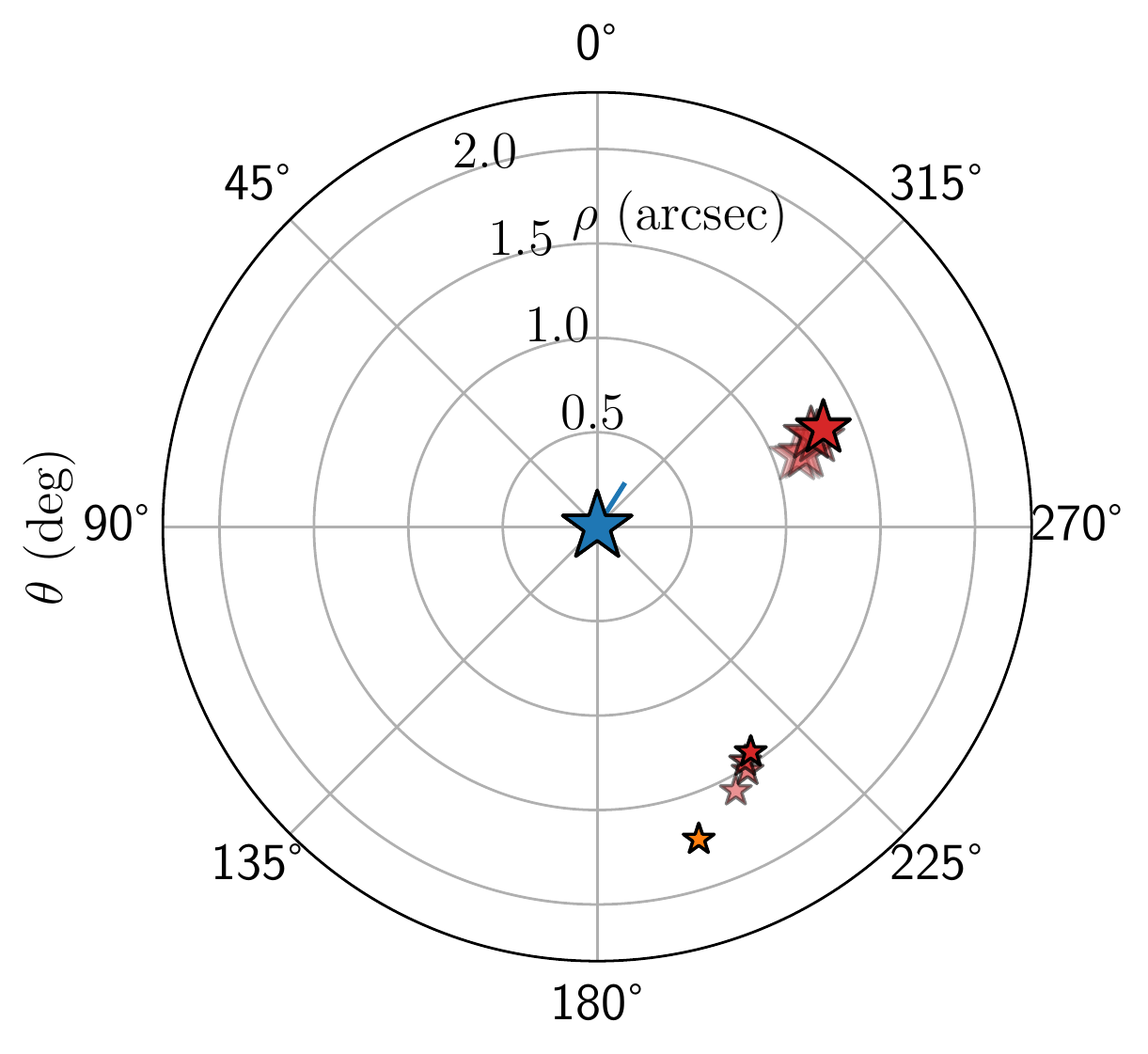}
    \caption{{\bf Sky positions of blended companions.} The blue star denotes TOI-1288, while red stars are the relative positions for the companions detected in the Spekle/AO images. Their sizes are scaled according to their relative brightness with the larger stars corresponding to $\Delta$Br-$\gamma=4.77$ and the smaller one corresponding to $\Delta$Br-$\gamma=5.88$. The transparent trails show how the companions move relative to TOI-1288 as a function of time with the opaque being the most recent position. The orange star is the relative position of the companion detected by \gaia, which is most likely the fainter companion. The blue line shows the proper motion of TOI-1288 over the course of 3.5 years.}
    \label{fig:rot_comp}
\end{figure}

\subsubsection{Are the companions bound?}\label{sec:bound}

In the following we will be referring to the brighter companion as star 1, and the fainter companion as star 2. To test whether these companions are bound, we study the positions of the host and the candidate companion in colour-magnitude diagrams (CMDs), loosely following the method outlined in \citet{Hirsch2017}. We used the measured photometry in the 832~nm and Br-$\gamma$ filters for star 1, and the \gaia $G$ and Br-$\gamma$ filters for star 2. In each case, we used the stellar parameters and uncertainties of the host ($\tau_\star$, [Fe/H], $\log g$, and $d$ from the SED fit in \tref{tab:spec_pars}) to generate a set of 1000 randomly sampled isochrones. For each filter pair, we determined a companion CMD position from the set of isochrones based on the $\Delta$-magnitude of the companion in each filter, and then calculated a weighted average of these measurements. This can then be compared to the observed CMD position of the companion as seen in \fref{fig:cmd}. For star 1, the observed and predicted positions agree to within 0.3$\sigma$, which could indicate that these objects are bound. This is further supported by their relative proximity on the sky. However, this could also be chance alignment for a background star with the right colour profile \citep{Hirsch2017}. For star 2, the observed and predicted CMD positions do not match, with a disagreement of \rev{3.5}$\sigma$. This strongly suggests that star 2 is a background star, and is not physically bound to the TOI-1288 system.

In \fref{fig:rot_comp} we show the relative positions of the companions detected in Speckle/AO and the one detected in \gaia. Evidently, the detected companions seem to be moving over the time span covered by the different observations in a similar direction, which is more or less opposite to the proper motion of TOI-1288. This clearly suggests that neither of the two companions are bound and are likely background stars. Finally, we note that the \gaia position is an average of different scans taken from July 2014 to May 2017 (for DR3) and might be less reliable. \rev{Furthermore, the reason that only the fainter companion was detected in the \gaia data could be that at an earlier epoch TOI-1288 and the brighter companion star were likely closer on the sky, and it would thus have been more difficult for \gaia to detect this companion. However, as seen in the speckle/AO observations, due to the proper motion of TOI-1288, the separation between TOI-1288 and this background star is increasing, meaning that it might be possible to detect it in future data releases.}

\rev{\subsection{High-resolution spectroscopy}
\label{sec:spec}

\subsubsection{FIES}

We performed high-resolution (R\,=\,$67\,000$) reconnaissance spectroscopy of TOI-1288 using the FIber-fed Echelle Spectrograph \citep[FIES;][]{Frandsen1999,Telting2014} mounted at the Nordic Optical Telescope \citep[NOT;][]{Djupvik2010} at Roque de los Muchachos Observatory, La Palma, Spain. The FIES spectra were extracted following \citet{Buchhave2010}, and stellar parameters were derived using the stellar parameter classification \citep[SPC;][]{Buchhave2012,Buchhave2014} tool. The resulting parameters are tabulated in \tref{tab:spec_pars}.

\subsubsection{HARPS-N}
\label{sec:HARPS-N}

We acquired 57 high-resolution (R\,=\,115\,000) spectra of TOI-1288 utilizing the High Accuracy Radial velocity Planetary Searcher for the Northern hemisphere \citep[HARPS-N;][]{art:cosentino2012} attached at the 3.58~m Telescopio Nazionale Galileo (TNG), also located at Roque de los Muchachos observatory. The spectra were collected between 19 November 2019 and 23 May 2022. We set the exposure time to 1200-2700\,s based on the sky conditions and scheduling constraints, which led to a median SNR of $\sim$60 per pixel at 550\,nm. We used the second fibre of the instrument to monitor the sky background.

The HARPS-N spectra were reduced and extracted using the dedicated Data Reduction Software \citep[DRS;][]{Lovis2007} available at the telescope. The DRS also provides the full width at half maximum (FWHM) and the bisector inverse slope (BIS) of the cross-correlation function (CCF), which was obtained by cross-correlating the observed \'echelle spectra against a G2 numerical mask. In this work, we used the Template-Enhanced Radial velocity Re-analysis Application \citep[TERRA;][]{Anglada2012} to extract precise RV measurements, along with additional activity indicators (namely, the H$\alpha$, S-index, and Na D indexes).

\subsubsection{HIRES}
\label{sec:hires}

We also gathered 28 spectra the High Resolution Echelle Spectrometer \citep[HIRES;][]{Vogt1994} mounted on the 10~m Keck-1 at the Keck Observatory, Hawai'i, USA. Observations were carried out between 10 December 2019 and 11 October 2021 with exposure times varying from 280-1000\,s depending on sky conditions, resulting in a median SNR of $\sim$72 near the spectral center of the image. The spectra were obtained with the iodine cell in the light path, and the RV extraction followed the standard HIRES forward-modelling pipeline \citep{Howard2010b}.}

\subsubsection{\rev{Periodogram Analysis}}
\label{sec:fup}

\begin{figure*}
    \centering
    \includegraphics[width=\textwidth]{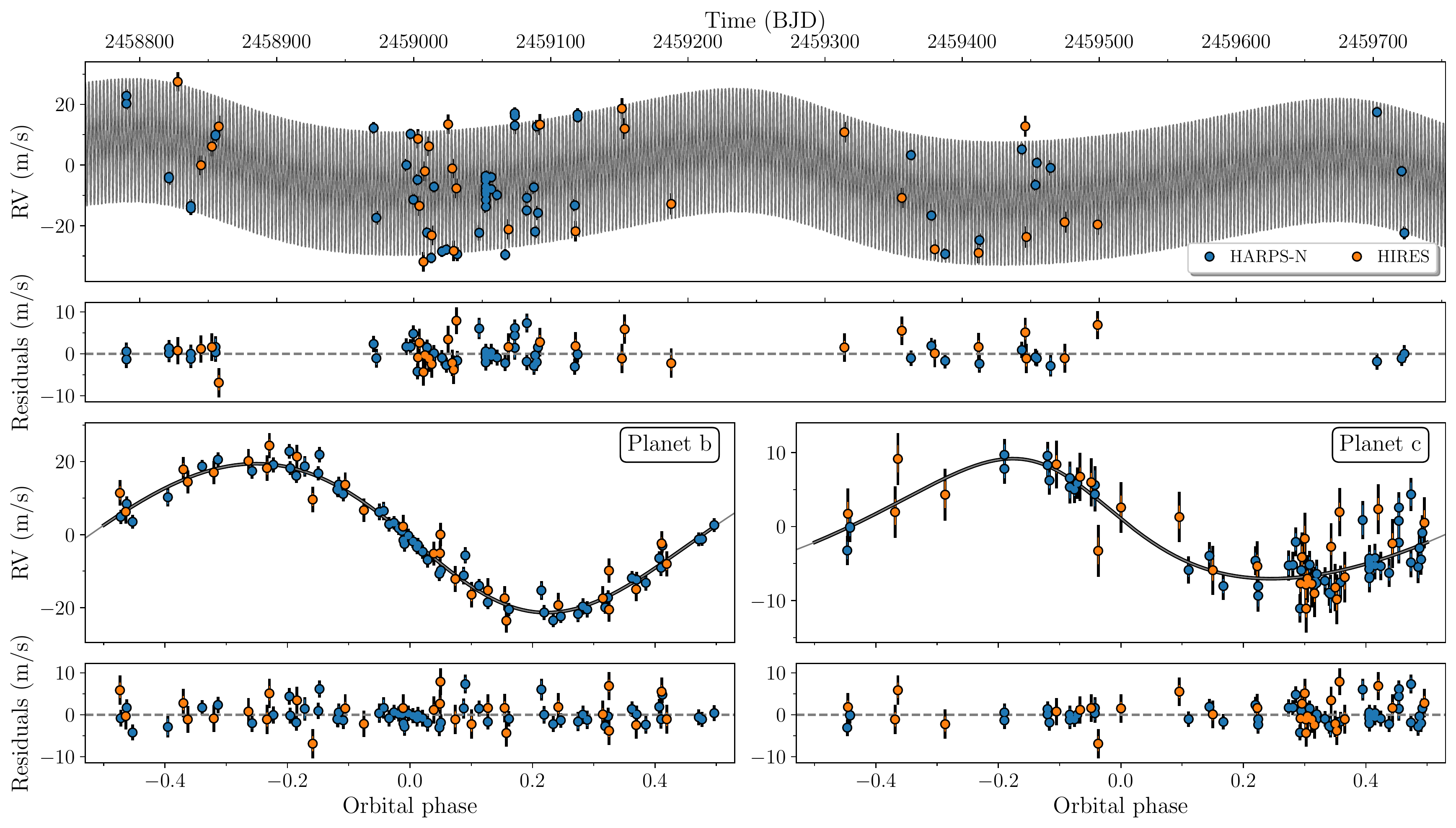}
    \caption{{\bf Radial velocities of TOI-1288.} {\it Top:} The HARPS-N (blue) and HIRES (orange) radial velocities as a function time. The grey model shows the combined signal for planet b and c. {\it Bottom left:} The radial velocities phased to the period planet b with the signal from planet c and the long-term trend subtracted with the best-fitting model overplotted. {\it Bottom right:} The radial velocities phased to the period of planet c with the signal from planet b and the long-term trend subtracted with the best-fitting model overplotted.}
    \label{fig:rvs}
\end{figure*}

\begin{figure}
    \centering
    \includegraphics[width=\columnwidth]{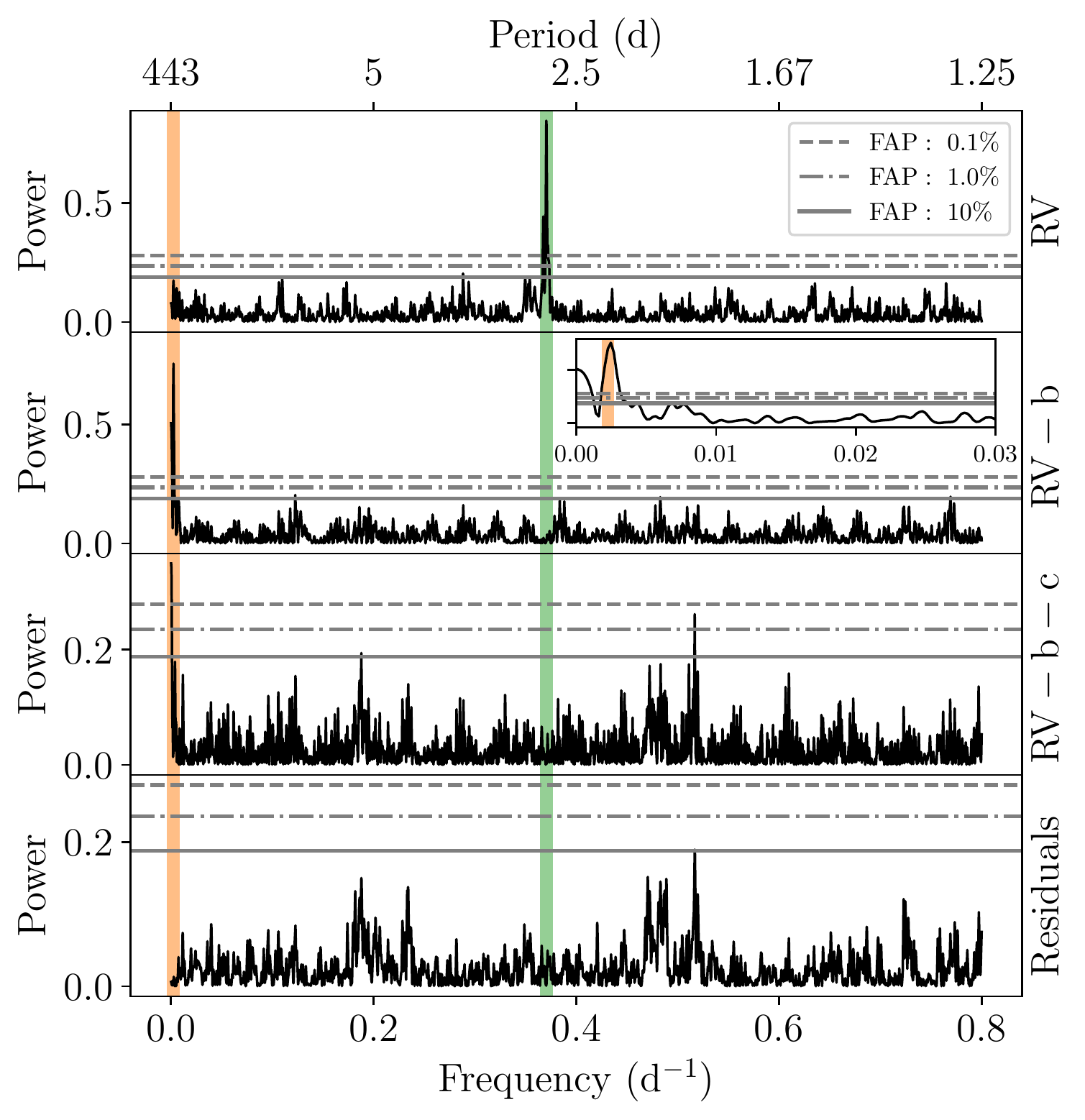}
    \caption{{\bf Generalised Lomb-Scargle diagram.} The GLS created from the RVs in \tref{tab:rvs}. {\it Top:} The GLS after subtracting the systemic velocities for HARPS-N and HIRES. The periods from \tref{tab:mcmc} for planet b (green) and planet c (orange) are shown as the vertical lines. The dashed, dashed-dotted, and solid horizontal lines are the 0.1\%, 1\%, and 10\% FAPs, respectively. {\it Upper middle:} The GLS after subtracting the signal from planet b.  The inset shows a close-up around the period of planet c. {\it Lower middle:} The GLS after subtracting both the signal from planet b and c. {\it Lower:} The GLS after subtracting both the signal from planet b, c, and the long-term trend. }
    \label{fig:gls}
\end{figure}

All the RVs are shown in \fref{fig:rvs} and tabulated in \tref{tab:rvs}. In \fref{fig:gls} we have calculated the generalised Lomb-Scargle \citep[GLS;][]{Lomb1976,Scargle1982} periodogram. Evidently, the $\sim$2.7~d transiting signal is also detected in the RVs, where a peak at this frequency clearly exceeds the false-alarm probabilities (FAPs; at 0.1\%, 1\%, and 10\%). We also see a significant peak at much lower frequencies with a period of around \Pcapp, which we ascribe to the presence of a further out companion. Seeing the \Pcapp-signal we searched the \tess light curve for additional transits  using the box least squares \citep[BLS;][]{Kovacs2002} algorithm after removing the transits from planet b ($\sim$2.7~d), but found no evidence for additional transiting signals.

\begin{figure}
    \centering
    \includegraphics[width=\columnwidth]{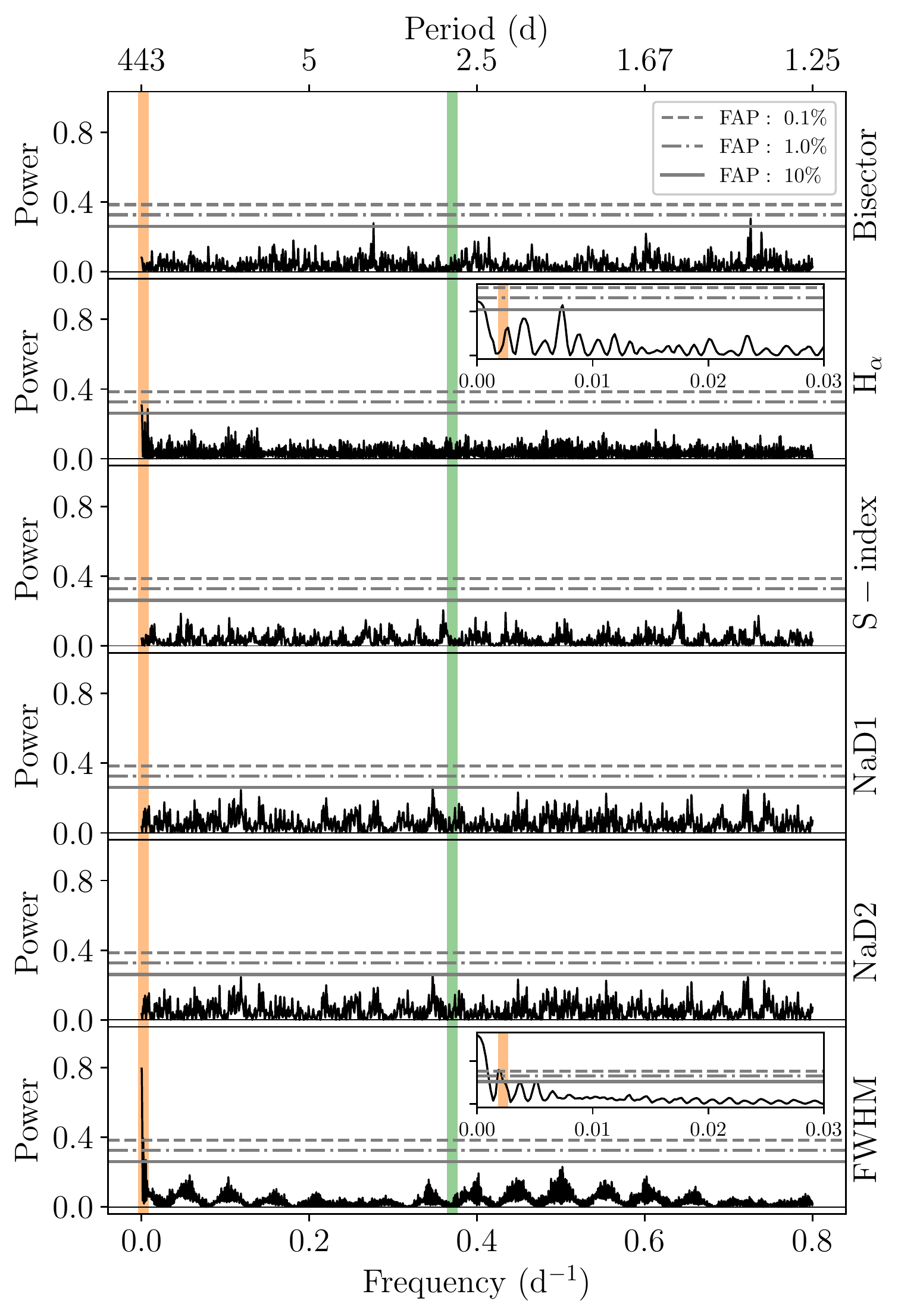}
    \caption{{\bf Generalised Lomb-Scargle diagram for activity indicators.} The GLS created from the activity indicators from the HARPS-N spectra. From top to bottom we show the GLS for the Bisector, H$_\alpha$, S-index, NaD1, NaD2, and full width half maximum (FWHM). Symbols have the same meaning as in \fref{fig:gls}.  }
    \label{fig:activity}
\end{figure}

We also detected another low frequency/long period peak in the GLS which seems to be a long-term trend in the RVs. We have furthermore created GLS periodograms for the activity indicators from the HARPS-N spectra shown in \fref{fig:activity}. Evidently, the star is inactive and the 2.7~d and \Pcapp do not coincide with any appreciable peak in these metrics, meaning that they are unlikely to come from stellar activity.

%\subsubsection{Reconnaissance}
%\label{sec:recon}

\subsubsection{Stellar modelling using SME and SED}
\label{sec:stellar}
In addition to the \rev{FIES} reconnaissance spectroscopy we also made use of our HARPS-N observations to derive stellar properties. We used co-added HARPS-N spectra with the software SME\footnote{\url{http://www.stsci.edu/~valenti/sme.html}.} \citep[Spectroscopy Made Easy;][]{Valenti1996, Piskunov2017}, a tool for fitting observations to synthetic spectra. A detailed description of the modelling can be found in \citet{Fridlund2017} and \citet{Persson2018}. For this star, we held the micro- and macro-turbulent velocities, $v_{\rm mic}$ and $v_{\rm mac}$, fixed in the modelling to 0.83~km~s$^{-1}$ \citep{Bruntt2010}, and 3.0~km~s$^{-1}$ \citep{Doyle2014}, respectively. The synthetic spectra were computed with the stellar atmosphere grid Atlas12 \citep{Kurucz2013}, and the atomic and molecular line data were taken from VALD\footnote{\url{http://vald.astro.uu.se}} \citep{Ryabchikova2015}. Our best model found an  effective temperature of $T_{\rm eff} = 5123 \pm 62$~K, an iron abundance of \mbox{[Fe/H]$ = +0.10\pm 0.11$}, a surface gravity of $\log g_\star = 4.23 \pm 0.09$, and a projected rotational velocity of $v \sin i_\star = 1.3 \pm 1.2$~km~s$^{-1}$. These results were checked with the empirical code SpecMatch-Emp \citep{Yee2017} and were found to agree within $1~\sigma$. 

Using the SME results as priors, we modelled the stellar radius with ARIADNE\footnote{\url{https://github.com/jvines/astroARIADNE}} \citep[][]{Vines2022} fitting broadband photometry to the spectral energy distribution (SED). The fitted bandpasses were the Johnson $B$ and $V$ magnitudes (APASS), $G G_{\rm BP} G_{\rm RP}$ (eDR3), $JHK_S$ magnitudes ({\it 2MASS}), {\it WISE} W1-W2, and the Gaia eDR3 parallax. The final radius was computed with Bayesian Model Averaging from the four fitted atmospheric models grids {\tt {Phoenix~v2}} \citep{Husser2013}, {\tt {BtSettl}} \citep{Allard2012}, \citet{Castelli2004}, and \citet{Kurucz1993} atmospheric model grids. The final stellar radius was found to be $1.010\pm 0.015$~$R_{\odot}$, and the stellar mass $0.895^{+0.042}_{-0.023}$~$M_{\odot}$ interpolated from the MIST \citep{Choi2016} isochrones. The stellar parameters are summarised in \tref{tab:spec_pars}.

\subsubsection{Stellar modelling using BASTA}\label{sec:basta}
As an independent measure for the stellar parameters we also modelled the star using the BAyesian STellar Algorithm\footnote{\url{https://basta.readthedocs.io/en/latest/index.html}} \citep[BASTA;][]{BASTA2015,BASTA2022}. We ran BASTA using the spectroscopic parameters from the SPC analysis ($T_{\rm eff}$, [Fe/H], $\log g$) as input along with the \gaia magnitudes ($G$, $B_{\rm P}$, $R_{\rm P}$) and parallax. BASTA's approach to fitting the magnitudes and parallax is described in Section 4.2.2 in \citet{BASTA2022}, where bolometric corrections are applied using the tables by \citet{Hidalgo2018}, and the reddening is calculated through the dust map by \citet{Green2019}. BASTA uses these values as constraints when fitting to a grid of BaSTI \citep[a Bag of Stellar Tracks and Isochrones;][]{Hidalgo2018} isochrones, where we opted for a science case that included both diffusion, convective core overshooting, and mass loss \citep[see Section 3.1 in][]{BASTA2022}. The resulting values are tabulated in \tref{tab:spec_pars} and are generally consistent with the other parameters, although BASTA found a slightly smaller stellar \rev{radius} as the fit seemed to prefer a slightly larger value for $\log g$ \rev{compared to the SME and SED fits}.
% as well as the 2MASS magnitudes from \tref{tab:star}

In the following we will be using stellar parameters coming from the SED. Therefore, derived quantities such as the planetary radius and masses will be calculated from the SED parameters.

\begin{table*}
    \centering
        \caption{ {\bf Stellar parameters for TOI-1288.} The stellar parameters from our spectral analyses and stellar modelling in \sref{sec:spec}, \sref{sec:stellar}, and \sref{sec:basta}. We also list the \gaia measurements.}
    \begin{threeparttable}    
    \begin{tabular}{c c c c c c c }
\toprule 
Parameter & Name & SME & SED & Spec-Match & SPC+BASTA\tnote{e} & {\it Gaia} DR3 \\ 
\midrule 
$T_\mathrm{eff}$ & Effective temperature (K) & $5123 \pm 62$ & 5225$^{+23}_{-27}$ & $5220 \pm 110$ & 5367$\pm$ 50 & $5300^{+20}_{-22}$ \\ 
 $\log g$ & Surface gravity & $4.23 \pm 0.09$ & $4.24 \pm 0.09$ & $4.36 \pm 0.12$ & $4.36\pm0.10$ & $4.447^{+0.010}_{-0.006}$ \\ 
 $\rm [Fe/H]$ & Iron abundance & $0.10 \pm 0.11$ & $0.07 \pm 0.09$ & $0.30 \pm 0.09 $& $0.18 \pm 0.08$ & $0.15^{+0.02}_{-0.03}$ \\ 
 $\rm [Ca/H]$ & Calcium abundance & $0.15 \pm 0.09$ & - & - & - & - \\ 
 $\rm [Na/H]$ & Sodium abundance & $0.25 \pm 0.12$ & - & - & - & - \\ 
 $v \sin i_\star$ & Projected rotation velocity (km~s$^{-1}$) & $1.3 \pm 1.2$ & - & - & <2 & - \\ 
 $\zeta$ & Macro-turbulence (km~s$^{-1}$) & 3.0\tnote{a} & - & - & - & - \\ 
 $\xi$ & Micro-turbulence (km~s$^{-1}$) & 0.83\tnote{b} & - & - & - & - \\ 
 $d$ & Distance (pc) & - & 114.7 $\pm$ 0.7 & - & $112.8^{+1.6}_{-1.4}$ & $114.677\pm0.013$ \\ 
 $R_\star$ & Stellar radius (R$_\odot$) & - & 1.010$^{+0.015}_{-0.014}$ & $1.09 \pm 0.18$ & $0.95_{-0.02}^{+0.03}$ & - \\ 
 $M_\star$\tnote{c} & Stellar mass (M$_\odot$) & - & 0.89$^{+0.04}_{-0.02}$ & 0.90 $\pm$ 0.08 & $0.91^{+0.04}_{-0.05}$ & - \\ 
 $M_\star$\tnote{d} & Stellar mass (M$_\odot$) & - & 0.65$^{+0.14}_{-0.13}$ & - & - & - \\ 
 $L_\star$ & Luminosity  (L$_\odot$) & - & $0.68 \pm 0.02$ & - & $0.65\pm0.03$ & - \\ 
 $A_V$ & $V$ band extinction & - & 0.014$^{+0.015}_{-0.009}$ & - & - & - \\ 
 $\tau_\star$ & Age (Gyr) & - & 12.1$^{+1.4}_{-3.1}$ & $10.05 \pm 0.17$ & $9.8^{+4.7}_{-3.8}$ & - \\ 
 \bottomrule
    \end{tabular}
\begin{tablenotes}
    \item[a] Relation from \citet{Doyle2014}.
    \item[b] Relation from \citet{Bruntt2010}.
    \item[c] SED estimate is from MIST isochrones.
    \item[d] SED estimate is from $\log g$ and $R_\star$.
    \item[e] $T_{\rm eff}$, $\log g$, [Fe/H], and $v \sin i_\star$ are from SPC. The rest have been derived using BASTA.
\end{tablenotes}
\end{threeparttable}
    \label{tab:spec_pars}
\end{table*}

\section{Analysis}
\label{sec:analysis}

In our modelling we included both planets, where only parameters for planet b are constrained by the photometry given we have not detected any transits of planet c. We modelled the transits using \texttt{batman}, where we accounted for the correlated noise in the light curve using Gaussian Process (GP) regression as implemented in \texttt{celerite} \citep{celerite}. We made use of the Matèrn-3/2 kernel, which is characterised by two hyper parameters: the amplitude, $A$, and the time scale, $\tau$. This model is shown at the bottom of \fref{fig:tess}.

In addition to the RV signals from planet b and c, we included a first-order acceleration parameter, $\dot{\gamma}$, to account for the long-term trend. Instead of stepping in $e$ and $\omega$, our Markov Chain Monte Carlo (MCMC) sampling was stepping in $\sqrt{e}\cos \omega$ and $\sqrt{e}\sin \omega$ for both planets. Furthermore, we were stepping in the sum of the limb-darkening coefficients, $q_1 + q_2$, while keeping the difference fixed. All stepping parameters and their priors are listed in \tref{tab:mcmc}.

As seen in \fref{fig:tpf} (see also \fref{fig:dss2} for a DSS2 image of the field) there are multiple companions in the \tess aperture mask. Therefore we added a dilution term in the MCMC, where we only included the contribution from all sources brighter than $\Delta G=5$, meaning that only the contribution from the south-eastern companion at a separation of $\sim$23~arcsec (\fref{fig:tpf}) was included. We thus did not consider the contribution from the (much closer) companions. The brightest of the two is found at $\Delta$Br-$\gamma=4.77\pm0.03$ and from our measurements in \tref{tab:speck} it is clear that both companions seem to be redder than TOI-1288, meaning that the differences in magnitude are even larger in the \tess passband.%The brightest of the two has a $\Delta$Br-$\gamma=4.77\pm0.03$ and 

The total flux as a function of time is thus $F(t)=(F_1(t)+F_2)/(F_1 + F_2)$, where $F_1(t)$ is the in-transit flux and $F_1$ is the out-of-transit flux for TOI-1288, and $F_2$ is the flux from the contaminating source at $\sim$23~arcsec. The flux from the contaminating source is then included as a fraction of TOI-1288, $F_1/F_2$, which in magnitude translates to $\Delta\mathrm{mag}=-2.5\log(F_2/F_1)$. As the \tess passband is very close to the \gaia $R_\mathrm{P}$ passband, we used the difference in this passband as a proxy for the difference between TOI-1288 and the 23~arcsec neighbour in the \tess passband. Thus we sampled the dilution as a Gaussian prior with $\Delta R_\mathrm{P} = \Delta \mathrm{TESS} = 4.41 \pm 0.02$. The photometric apertures from the ground-based facilities are small enough (\tref{tab:transitfollowup}) so that this source does not contaminate those light curves. As such no dilution factors were included for these.

We sampled the posteriors for the transit and orbital parameters using MCMC sampling utilising the \texttt{emcee} package \citep{emcee}.
Our likelihood function is defined as
\begin{equation}
    \label{equ:likelihood}
    \log \mathcal{L} =-0.5 \sum_{i=1}^{N} \left [ \frac{(O_i - C_i)^2}{\sigma_i^2} + \log 2 \pi \sigma_i^2 \right] \, ,
\end{equation}
where $N$ indicates the total number of data points from photometry and RVs. $C_i$ represents the model corresponding to the observed data point $O_i$. $\sigma_i$ represents the uncertainty for the $i$th datum, where we add a jitter term in quadrature and a penalty in the likelihood for the RVs. To our likelihood in \eref{equ:likelihood} we add our priors $\sum_{j=1}^{M} \log \mathcal{P}_{j}$, $\mathcal{P}_j$ being the prior on the $j$th parameter, and this sum constitutes the total probability.

\begin{table*}
    \centering
    \caption{{\bf MCMC results.} The median and high posterior density at a confidence level of 0.68. Subscripts b and c denote parameters for planet b and c, respectively. $\mathcal{U}$ denotes that a uniform prior was applied during the run.
    %, while $\mathcal{N}(\mu,\sigma)$ is a Gaussian prior with mean $\mu$ and width $\sigma$. $\mathcal{F}(c)$ means that this quantity was fixed at $c$ during the run. 
    }
    \begin{threeparttable}    

    \begin{tabular}{l l c c}
\toprule 
Parameter & Name & Prior & Value \\ 
\midrule 
\multicolumn{4}{c}{Stepping parameters} \\ 
\midrule 
$P_\mathrm{b}$ & Period (days) & $\mathcal{U}$ & $2.699835^{+0.000004}_{-0.000003}$ \\ 
$T_\mathrm{0,b}$ & Mid-transit time (BTJD) & $\mathcal{U}$ & $1712.3587 \pm 0.0002$ \\ 
$(R_\mathrm{p}/R_\star)_\mathrm{b}$ & Planet-to-star radius ratio & $\mathcal{U}$ & $0.0476 \pm 0.0005$ \\ 
$(a/R_\star)_\mathrm{b}$ & Semi-major axis to star radius ratio & $\mathcal{U}$ & $8.5 \pm 0.4$ \\ 
$K_\mathrm{b}$ & Velocity semi-amplitude (m s$^{-1}$) & $\mathcal{U}$ & $20.7^{+0.4}_{-0.5}$ \\ 
$\cos i_\mathrm{b}$ & Cosine of inclination & $\mathcal{U}$ & $0.030^{+0.012}_{-0.030}$ \\ 
$(\sqrt{e} \cos \omega)_\mathrm{b}$ &   & $\mathcal{U}$ & $-0.19^{+0.03}_{-0.04}$ \\ 
$(\sqrt{e} \sin \omega)_\mathrm{b}$ &   & $\mathcal{U}$ & $0.16^{+0.07}_{-0.06}$ \\ 
$P_\mathrm{c}$ & Period (days) & $\mathcal{U}$ & $443^{+11}_{-13}$ \\ 
$T_\mathrm{0,c}$ & Mid-transit time (BTJD) & $\mathcal{U}$ & $1883^{+12}_{-14}$ \\ 
$K_\mathrm{c}$ & Velocity semi-amplitude (m s$^{-1}$) & $\mathcal{U}$ & $7.6^{+0.5}_{-0.6}$ \\ 
$(\sqrt{e} \cos \omega)_\mathrm{c}$ &   & $\mathcal{U}$ & $0.15^{+0.19}_{-0.15}$ \\ 
$(\sqrt{e} \sin \omega)_\mathrm{c}$ &   & $\mathcal{U}$ & $0.28^{+0.14}_{-0.13}$ \\ 
$\gamma_\mathrm{HARPS-N}$ & Systemic velocity HARPS-N (m s$^{-1}$) & $\mathcal{U}$ & $7.7^{+0.8}_{-0.7}$ \\ 
$\sigma_\mathrm{HARPS-N}$ & Jitter HARPS-N (m s$^{-1}$) & $\mathcal{U}$ & $1.9 \pm 0.3$ \\ 
$\gamma_\mathrm{HIRES}$ & Systemic velocity HIRES (m s$^{-1}$) & $\mathcal{U}$ & $6.2^{+0.9}_{-1.0}$ \\ 
$\sigma_\mathrm{HIRES}$ & Jitter HIRES (m s$^{-1}$) & $\mathcal{U}$ & $3.4 \pm 0.6$ \\ 
$\dot{\gamma}$ & Linear trend (m s$^{-1}$ d$^{-1}$) & $\mathcal{U}$ & $-0.0088 \pm 0.0017$ \\ 
\midrule 
\multicolumn{4}{c}{Derived parameters} \\ 
\midrule 
$e_\mathrm{b}$ & Eccentricity & - & $0.064^{+0.014}_{-0.015}$ \\ 
$\omega_\mathrm{b}$ & Argument of periastron ($^\circ$) & - & $139^{+13}_{-17}$ \\ 
$i_\mathrm{b}$ & Inclination ($^\circ$) & - & $88.3^{+1.7}_{-0.7}$ \\ 
$b_\mathrm{b}$ & Impact parameter & - & $0.26^{+0.10}_{-0.24}$ \\ 
$e_\mathrm{c}$ & Eccentricity & - & $0.13^{+0.07}_{-0.09}$ \\ 
$\omega_\mathrm{c}$ & Argument of periastron ($^\circ$) & - & $63^{+30}_{-33}$ \\ 
$T_\mathrm{14,b}$ & Transit duration (hours) & - &$ 2.37_{-0.03}^{+0.05} $ \\ 
\midrule 
\multicolumn{4}{c}{Physical parameters \tnote{\textdagger}} \\ 
\midrule 
$T_\mathrm{eq,b}$\tnote{$\chi$} & Equilibrium temperature (K) & - &$ 1266 \pm 27 $ \\ 
$R_\mathrm{p,b}$ & Planet radius (R$_\oplus$) & - &$ 5.24 \pm 0.09 $ \\ 
$M_\mathrm{p,b}$ & Planet mass (M$_\oplus$) & - &$ 42 \pm 3 $ \\ 
$\rho_\mathrm{p,b}$ & Planet density (g~cm$^{-3}$) & - &$ 1.3 \pm 0.5 $ \\ 
$(M_\mathrm{p} \sin i)_\mathrm{c}$ & Lower value for planet mass (M$_\oplus$) & - &$ 84 \pm 7 $ \\ 
\bottomrule

    \end{tabular}
\begin{tablenotes}
    \item[*] Barycentric TESS Julian Date (BTJD) is defined as BJD-2457000.0, BJD being the Barycentric Julian Date
    \item[\textdagger] From the SED stellar parameters in \tref{tab:spec_pars}.
    \item[$\chi$] Following \citet{Kempton2018}.
\end{tablenotes}
\end{threeparttable}    
    \label{tab:mcmc}
\end{table*}

\section{Results}\label{sec:results}
\begin{figure}
    \centering
    \includegraphics[width=\columnwidth]{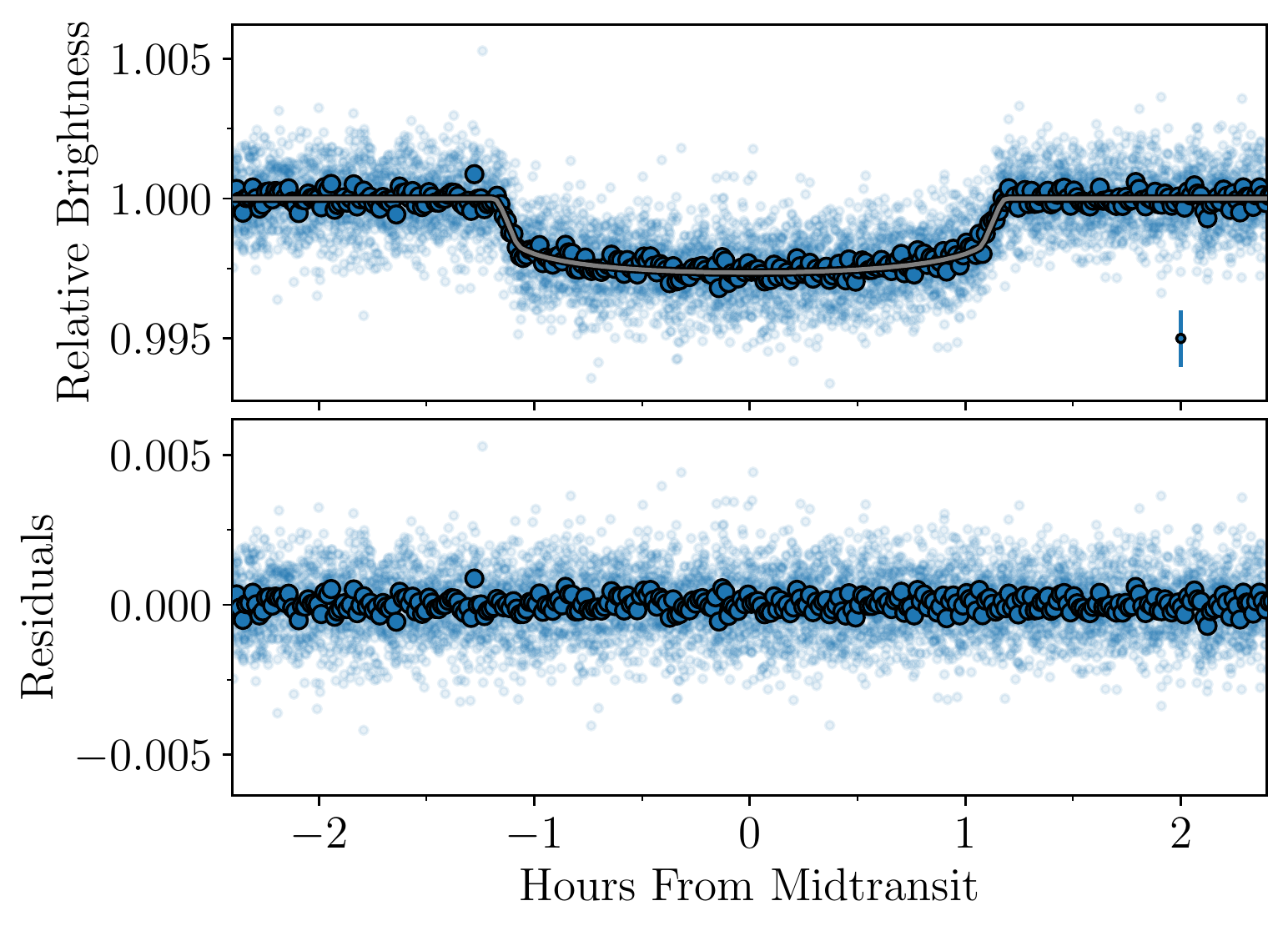}
    \caption{{\bf \tess light curve of TOI-1288~b.} The GP detrended \tess data from \fref{fig:tess} showing the phase folded transits of planet b. We show the data binned in larger, solid points, while the unbinned data are shown smaller, more transparent points. The datum with errorbar is not an actual measurement, but illustrates the median of the uncertainties of all data. The grey line is the best-fitting model.}
    \label{fig:lcbTESS}
\end{figure}

In \fref{fig:lcbTESS} we show the \tess light curve phase-folded on the transits of planet b along with the best-fitting model. Light curves from all photometric observations can be found in \fref{fig:lcb}. We find a planet-to-star radius ratio of \rprsb, which given the stellar radius from the SED analysis in \tref{tab:spec_pars} yields a radius of \rpb. With a period of just \Pb, TOI-1288~b is thus a hot super-Neptune.

Shown in \fref{fig:rvs} are the best-fitting models for the radial velocities for both planet b and c. This 2-planet model is heavily favoured over a 1-planet model according to the Bayesian information criterion (BIC, $\Delta$BIC$=104$). To get a measure of the mass for both planets we use the relation 

\begin{equation}
    M_{\rm p} \sin i = \frac{K \sqrt{1 - e^2}}{28.4~\mathrm{m}~\mathrm{s}^{-1}} \left ( \frac{P}{1~\mathrm{yr}} \right)^{1/3} \left ( \frac{M_\star}{\mathrm{M}_\odot} \right)^{2/3}  \, ,
    \label{eq:mass}
\end{equation}
where we can only get a lower limit for the mass of planet c as we do not know the inclination. For planet b we find a mass of \mpb, which combined with the radius yields a bulk density of \rhob. For planet c we find a lower limit for the mass of \mpc.

For the long-term trend we have found a value for $\dot{\gamma}=-0.0086\pm0.0019$~m~s$^{-1}$~d$^{-1}$. This first-order acceleration parameter constitutes a lower limit for the semi-amplitude through $(t_{\rm f}-t_{\rm i})\times \dot{\gamma}/2$ with $t_{\rm f}$ and $t_{\rm i}$ being the final and first timestamps. Following the Monte Carlo approach in \citet{Kane2019} \citep[see also][]{Pepper2020}, we used our measured value for $\dot{\gamma}$ to calculate the lower limit for the companion inducing this long-term trend as a function of orbital separation. Namely, we solved 
\begin{equation}
    K \leq \sqrt{\frac{G}{a_{\rm B}(1-e_{\rm B}^2)}}\frac{M_{\rm B} \sin i_{\rm B}}{\sqrt{M_{\rm B} + M_\star}}
\end{equation}
for M$_{\rm B}$ at each $a_{\rm B}$ with $e_{\rm B}$ being drawn from a $\beta$-distribution and $\cos i_{\rm B}$ from a uniform distribution. 

\begin{figure}
    \centering
    \includegraphics[width=\columnwidth]{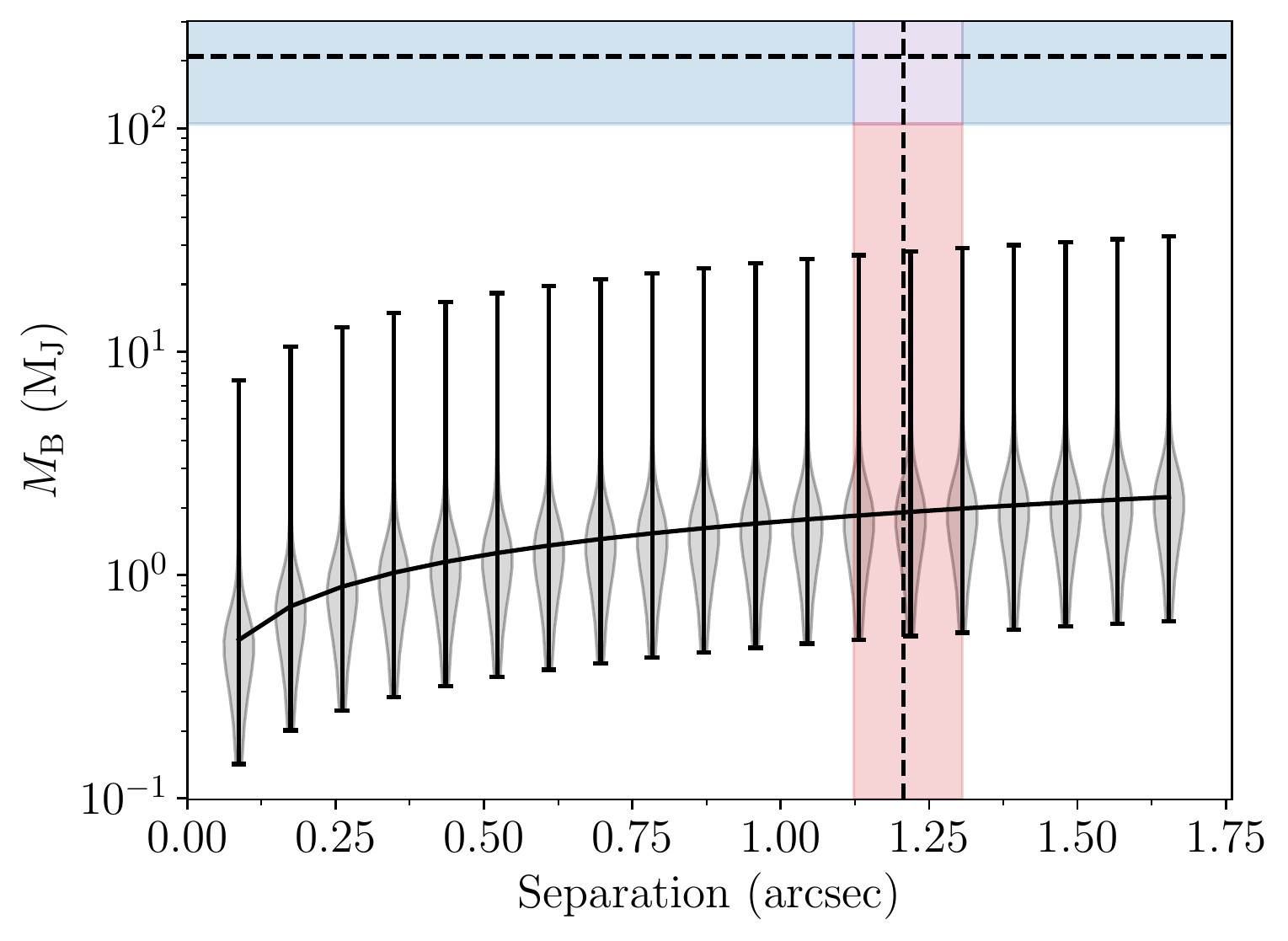}
    \caption{{\bf Lower mass limit for an additional bound companion.} The violinplot shows the resulting distribution for for the mass of the companion for a given separation (here converted to sky-projected separation using the distance from \tref{tab:star}). The solid black curve is the median mass for each separation. The vertical red band spans the range of the speckle and AO measurements for the separation of star 1 (\tref{tab:speck}), while the dashed vertical line is the median of these. The horizontal blue band spans the range from 0.1~M$_\odot$ to 0.3~M$_\odot$ with 0.2~M$_\odot$ shown with the dashed line.}
    \label{fig:lower}
\end{figure}

In \fref{fig:lower} we show the resulting distributions for each orbital separation, here converted to a sky-projected separation. We furthermore show the observed position of the brightest of the two companions, star 1, detected in speckle and AO, {\it if} it were bound to TOI-1288. From our analysis in \sref{sec:bound} and its position in the CMD in \fref{fig:cmd} this companion would most likely have been an M-dwarf with a mass of around $0.2$~M$_\odot$. While this is a lower limit for the mass and could be consistent with the mass we have estimated for star 1, the median is around two orders of magnitude lower at the position for star 1. We should thus in most cases have detected a much more significant drift, if it were due to star 1. Therefore, it seems more likely that the drift we are seeing is coming from another planetary companion. 

\rev{Finally, we note that the \gaia Renormalised Unit Weight Error (RUWE) statistic for TOI-1288 is 1.17. For a good single-star fit one would expect it to be around 1, whereas a value of $\gtrapprox1.4$ could suggest that the source is non-single or otherwise problematic for the astrometric solution. The slight departure could be because the \gaia astrometry is seeing the orbital motion induced by this long-term RV companion.}

\section{Discussion}
\label{sec:disc}
\subsection{Location in the Neptunian desert}

\begin{figure*}
    \centering
    \includegraphics[width=\textwidth]{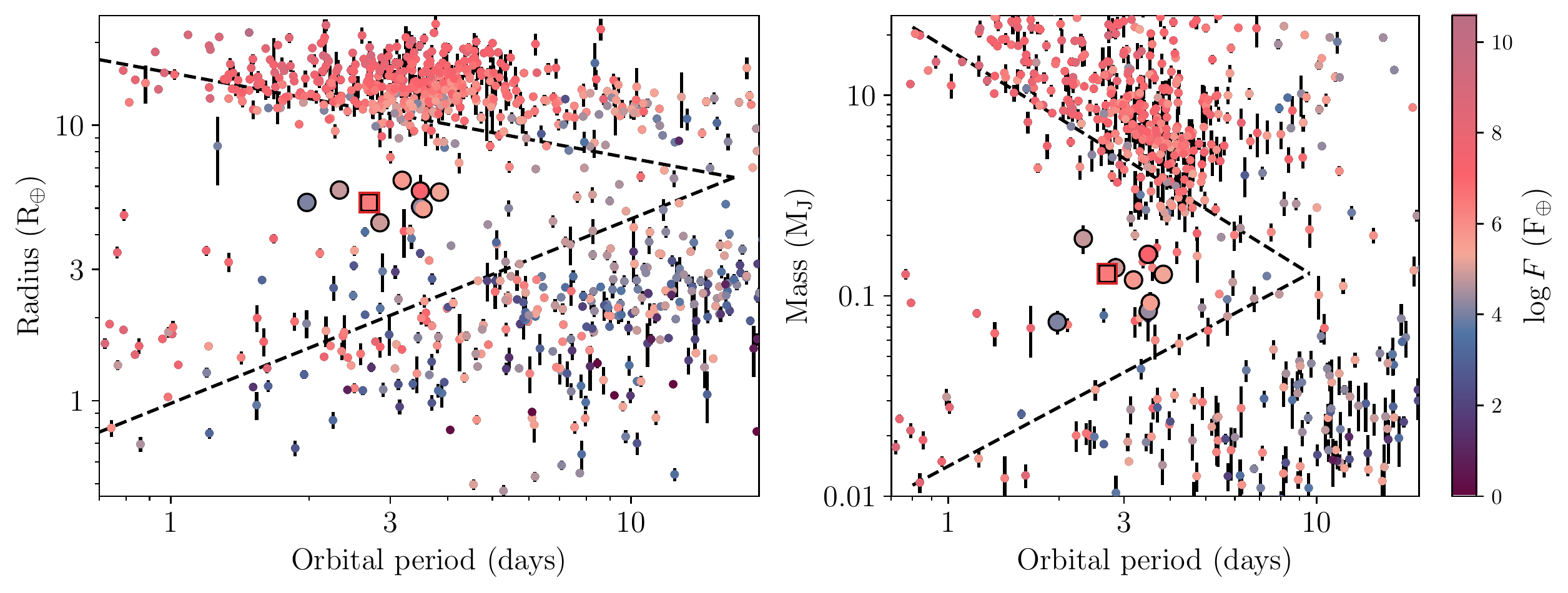}
    \caption{{\bf The hot Neptunian desert} reported in \citet{Mazeh2016} shown as dashed lines. Planets (as of September 2022) from the \texttt{TEPCat} catalogue of "well-studied transiting planets" \citep[][ \url{https://www.astro.keele.ac.uk/jkt/tepcat/allplanets-noerr.html}]{Southworth2011} with uncertainties smaller than 30\% in radius (left) and mass (right). The points are colour coded according to the incident flux, which is truncated at $F=1$~F$_\oplus$. TOI-1288~b is shown as the large square with a red outline. The large circles denote the closest eight planets to TOI-1288~b in the radius-period parameter space, with their position highlighted in the mass-radius diagram as well.}
    \label{fig:desert}
\end{figure*}

We have found TOI-1288~b to be a hot super-Neptune with an equilibrium temperature of \teq \citep[estimated from][assuming zero albedo and full day-night heat redistribution]{Kempton2018}. In \fref{fig:desert} we plot the radius (left, in Earth radii) and mass (right, in Jupiter masses) of TOI-1288~b as functions of orbital period. Evidently, TOI-1288~b falls right in the hot Neptunian desert reported by \citet{Mazeh2016}. \citet{Mazeh2016} mention two processes that could account for the upper boundary. Firstly, if the planet had migrated through the disk, then stopped at the upper boundary of the desert due to a decrease in density in the disk as it moves inwards, the inner radius of the disk might be related to its mass and consequently the planetary mass. Therefore, the central hole in the disk would be smaller in a more massive disk, and hence allow for a more massive planet. Alternatively, the atmosphere of a planet moving horizontally in the diagram, i.e., migrating, might be stripped of its atmosphere due to the stellar irradiation, resulting in a smaller, lower mass planet. 

In \citet{Vissapragada2022} the upper boundary of the desert was investigated by looking at the metastable helium feature in the atmospheres of the planets, which could be a tracer for any outflows. They found that this upper boundary is stable against photoevaporation, meaning that a different mechanism must be responsible for tracing out this upper edge. This is in-line with the findings of \citet{Owen2018} in which they argue that if photoevaporation is responsible for the upper boundary, we should see a lot of sub-Jovian mass planets in the mass-period plane at very short periods, which we do not. Rather they argue that the upper boundary is caused by high-eccentricity migration.

On the other hand, \citet{Owen2018} do find that the lower boundary could be explained by photoevaporation. This photoevaporation which leaves behind a rocky core has furthermore been used to explain the dearth of {\it hot super-Earths} \citep[e.g.,][]{Sanchis2014,Lundkvist2016}. An alternative explanation for the lower boundary of the desert could be that as the separation increases, so does the Hill sphere of the planetesimal, the orbital path, and the dust-to-gas ratio, meaning that the core mass is increased towards the end of the first stage of formation. This would then result in more massive planets at larger separations \citep{Mazeh2016}.

What is clear from \fref{fig:desert} is that the upper boundary is much more well-defined than the lower boundary. However, even if the lower boundary would be at larger radii, TOI-1288~b is still found in a very deserted area. In \fref{fig:desert} we have highlighted \rev{eight} planets that are the closest to TOI-1288~b in the radius-period (distance here measured in units of $(\mathrm{R}_\oplus^2 + \mathrm{days}^2)^{1/2}$) plane; {\it Kepler}-101~b \citep{Bonomo2014}, HATS-7~b \citep{Bakos2015}, TOI-532~b \citep{Kanodia2021}, TOI-674~b \citep{Murgas2021}, TOI-1728~b \citep{Kanodia2020}, NGTS-14Ab \citep{Smith2021}, WASP-156~b \citep{Demangeon2018}, and K2-55~b \citep{Crossfield2016}. Some key parameters \citep[][from \url{https://www.astro.keele.ac.uk/jkt/tepcat/allplanets-noerr.html}]{Southworth2011} for these systems are summarised in \tref{tab:comp} along our parameters for TOI-1288~b. 

Obviously, the planets are similar in terms of period and radius, but they also have quite similar masses, and thus densities. The most striking difference in \fref{fig:desert} is the insolation, which is dictated by the spectral type ($T_\mathrm{eff}$) of the stellar host. In this context it is worth noting that the overabundance of large planets with high insolation compared to at smaller radii in \fref{fig:desert}, merely reflects that it is easier to detect a larger planet around a larger (hotter) star. This is apparent from \fref{fig:desert_teff} and also what is seen in \citet{Szabo2019}.

A clustering of Neptune-sized planets with equilibrium temperatures of around 2000~K has been reported in \citet{Persson2022}, which begs the question whether there could be an island of stability in the desert. However, this might also be a selection effect, and more planets in this parameter space are needed to establish this. It is an intriguing idea, and if an island of stability could exist for these slightly smaller planets on more irradiated orbits, maybe a similar island exist for TOI-1288~b and its neighbours, who are slightly bigger and less irradiated. It could also be a strip of pseudo stability in the desert, or it might just reflect the aforementioned less well-defined lower boundary of the desert.

\begin{table}
    \centering
    
    \caption{{\bf Closest radius-period neighbours. The eight planets closest to TOI-1288~b in terms of radius and period (with distance in units of $(\mathrm{R}_\oplus^2 + \mathrm{days}^2)^{1/2}$).} 
    }
    \begin{threeparttable}    
    
    \begin{tabular}{c c c c c c c}
    \toprule

         & $P$ & $F$\tnote{\textdagger} & $R_{\rm p}$ & $M_{\rm p}$ & $\rho_{\rm p}$ & SpT \\ 
          & (d) & (F$_\oplus$) & (R$_\oplus$) & (M$_\oplus$) & ($\rho_\oplus$) & \\
         \midrule
         TOI-1288~b & 2.6998 & 630 & 5.6 & 41 & 0.24 & G \\
         TOI-532~b &  2.327 & 119 & 5.8 & 61 & 0.31 & M \\
         TOI-674~b & 1.977 & 57 & 5.3 & 24 &  0.17 & M \\
         {\it Kepler}-101~b & 3.488 & 1260 & 5.8 & 51 & 0.26 & G \\
         HATS-7~b & 3.185 & 288 & 6.3 & 38 & 0.15 & K \\
         TOI-1728~b & 3.492 &  72 & 5.1 & 27 & 0.21 & M \\
         NGTS-14Ab & 3.536 & 240 & 5.0 & 29 & 0.25 & K \\ 
         WASP-156~b & 3.836 & 186 & 5.7 & 41 & 0.24 & K \\
         K2-55~b & 2.849 & 130 & 4.4 & 44 & 0.5 & K \\
    \bottomrule         
    \end{tabular}
\begin{tablenotes}
    \item[\textdagger] From $ (\rho_\star/\rho_\odot)^{-2/3} (P/1~\mathrm{yr})^{-4/3} (T_\mathrm{eff}/5777~\mathrm{K})^4$.
    %\item[$\chi$] Following \citet{Kempton2018}.
\end{tablenotes}
\end{threeparttable}    
    \label{tab:comp}
\end{table}

\subsection{Internal structure and atmosphere}
\label{sec:inter}
\begin{figure}
    \centering
    \includegraphics[width=\columnwidth]{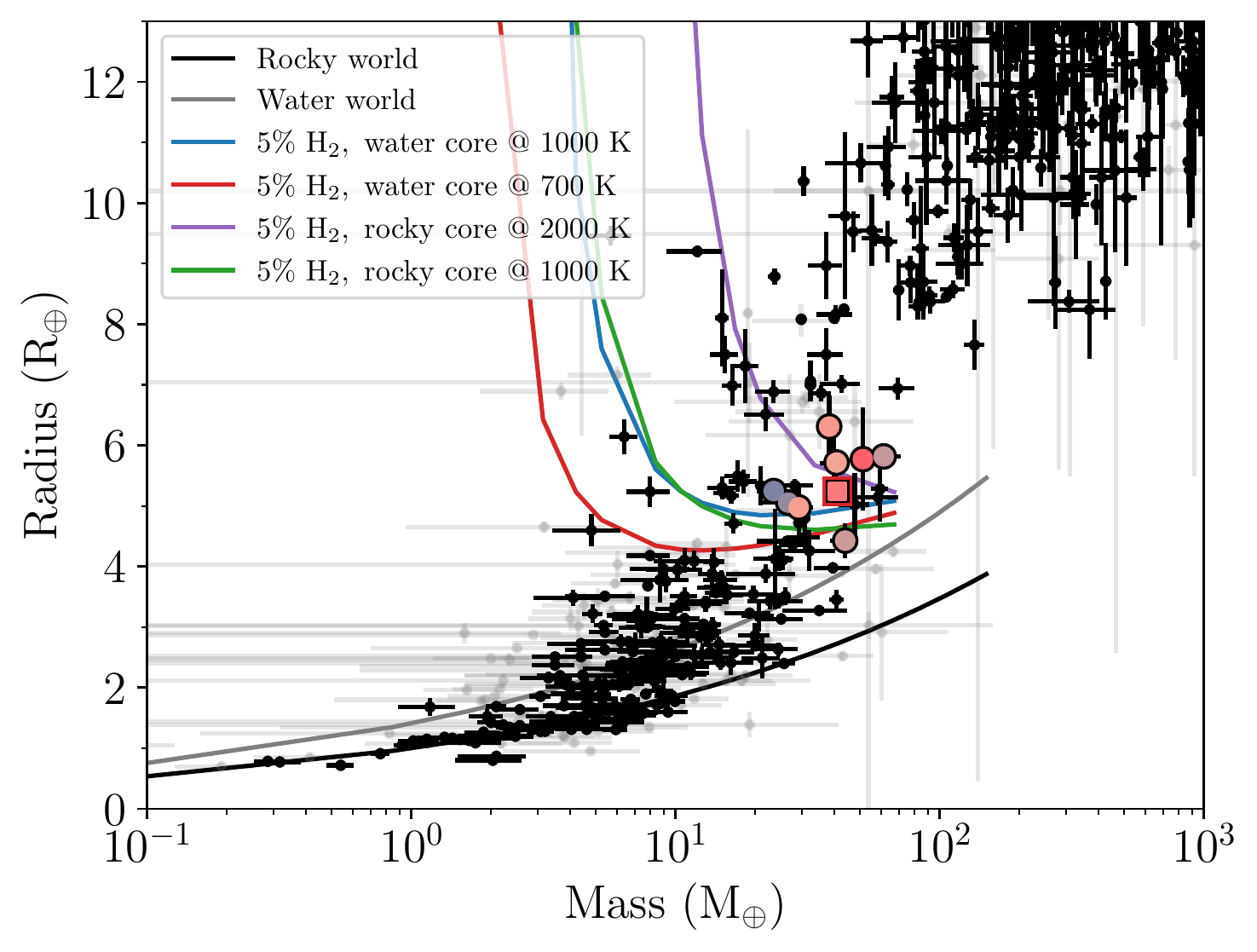}
    \caption{{\bf Mass-radius diagram.} Planets from the same catalogue as in \fref{fig:desert}, but now for planets with uncertainties on both mass and radius of less (more) than 30\% shown as black (grey) dots. Solid lines are composition models from \citet{Zeng2016,Zeng2019}. TOI-1288~b is again shown as the large (coloured) square with the similar ($R_\mathrm{p},P$) planets shown with large circles.}
    \label{fig:r_vs_m}
\end{figure}

In the mass-radius diagram in \fref{fig:r_vs_m} we compare TOI-1288~b to models with different compositions. The models are taken from \citet{Zeng2016,Zeng2019}. Evidently, TOI-1288~b can be described as a rocky core with a gaseous envelope at high irradiation. Probing the atmosphere of the planet through transmission spectroscopy could naturaully help reveal atmospheric features, but can also provide valuable constraints on the internal structure.

We observed a transit of planet b on the night 2020 June 11 using the HARPS-N spectrograph. The RVs from this night can be seen around orbital phase 0.0 in the lower left panel of \fref{fig:rvs}, but due to the slow rotation of the star ($v \sin i_\star=1.3\pm1.2$~km~s$^{-1}$) we do not see the Rossiter-McLaughlin \citep[RM;][]{Rossiter1924,McLaughlin1924} effect. For an aligned configuration a decent approximation for the amplitude is given by $0.7\sqrt{1-b^2}(R_{\rm P}/R_\star)^2v \sin i_\star$, which comes out to just shy of 2~m~s$^{-1}$ for TOI-1288~b. 

We nonetheless ran an MCMC where we included the RM effect \citep[using the code by][]{Hirano2011}. We excluded the photometry and instead applied Gaussian priors to $P_{\rm b}$, $T_{\rm 0, b}$, $(R_{\rm p}/R_\star)_{\rm b}$, $(a/R_\star)_{\rm b}$, and $i_{\rm b}$ using the values in \tref{tab:mcmc} and for $v \sin i_\star$ from the SME value in \tref{tab:spec_pars}. We also applied Gaussian priors to the macro- and micro-turbulence as well as the sum of the limb-darkening coefficients \citep[values estimated from][respectively]{Doyle2014,Bruntt2010,Claret2011}, while applying a uniform prior to the sky-projected obliquity, $\lambda_{\rm b}$. The rest followed the same approach as the run in \sref{sec:analysis}. The resulting value for the projected obliquity was $\lambda_{\rm b}=70^{+110}_{-100}$$^\circ$, meaning that it is unconstrained.    

Following \citet{Kempton2018} we can calculate the transmission spectroscopic metric (TSM) to assess the feasibility of transmission spectroscopy for TOI-1288~b. The TSM is given by

\begin{equation}
    \mathrm{TSM} = H \times \frac{R_{\rm p}^{3} T_{\rm eq}}{M_{\rm p} R_\star^2} \times 10^{-m_J/5} \, ,
    \label{eq:tsm}
\end{equation}
where $m_J$ is the apparent magnitude of the host in the $J$ band and $H$ is a scale factor related to the size of the planet. For TOI-1288~b $H$ is 1.15, while the planet, stellar, and system parameters are listed in \tref{tab:mcmc}, \tref{tab:spec_pars} (SED), and \tref{tab:star}, respectively. This yields a TSM of $\sim87$, which is just below the suggested cutoff for follow-up efforts in \citet{Kempton2018}.

\begin{figure}
    \centering
    \includegraphics[width=\columnwidth]{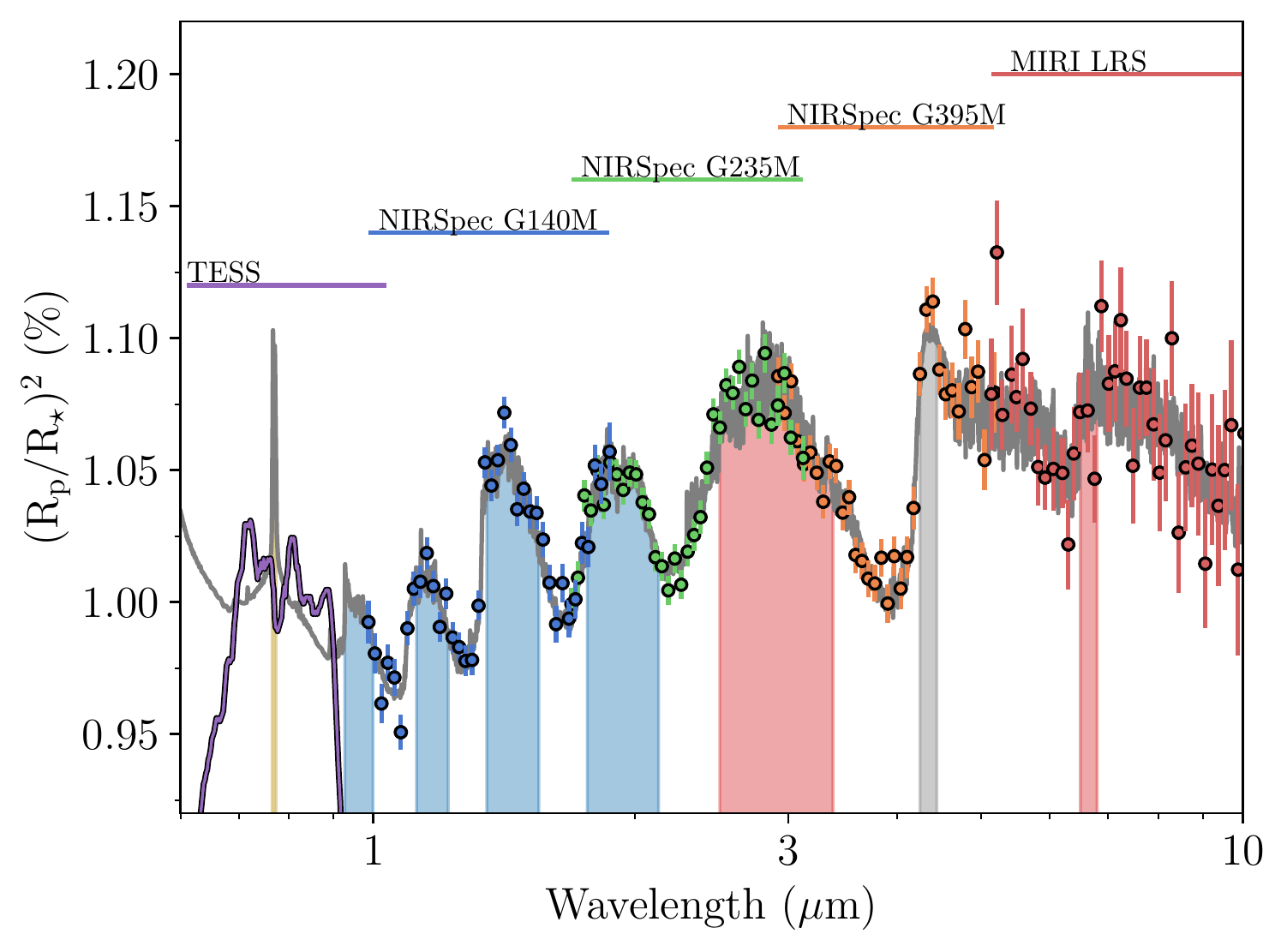}
    \caption{{\bf Simulated JWST observations.} A simulated transmission spectrum of TOI-1288~b in grey using petitRADTRANS. The coloured error bars are simulated JWST data from PandExo of different instruments with their wavelength coverage shown by the horizontal coloured lines and with the names of the instrument shown above. We also show the \tess transmission curve in purple. Some atomic and molecular species are highlighted in the coloured areas with K, H$_2$O, CH$_4$, and CO$_2$ shown with yellow, blue, red, and grey, respectively. }
    \label{fig:jdub}
\end{figure}

While -- according to this metric -- TOI-1288~b is not a high priority target for JWST \citep{Gardner2006}, we still investigate what JWST might be able to detect if it were to do transmission spectroscopy. We simulated the spectrum of TOI-1288~b using petitRADTRANS \citep{Molliere2019,Molliere2020} assuming a cloud-free, isothermal model at 1266~K. We used PandExo \citep{Batalha2017} to simulate the JWST data for four different instruments. For each we assumed a total of 4 transits with a 4~hr baseline each. The resulting spectrum is shown in \fref{fig:jdub}. For this most likely quite optimistic scenario, JWST should be able to detect several molecular species, such as H$_2$O, CH$_4$, and CO$_2$ (if present).

\subsection{Outer companions}
\label{sec:comp}
According to the Web \tess Viewing Tool\footnote{\url{https://heasarc.gsfc.nasa.gov/cgi-bin/tess/webtess/wtv.py}}, TOI-1288 is (at the time of writing) being re-observed in Sectors 56-58 (beginning in September 2022 and ending in November 2022). These additional sectors should help refine the transit parameters of planet b. While our current estimate for the period and ephemeris of planet c suggest a transit occurred (of course, depending on the inclination) July 2022, the uncertainties are rather large, so it is worthwhile to be on the look out for a potential transit of planet c. % or any other potential planet in the system.

\citet{Zhu2018} and \citet{Bryan2019} found an excess of cold Jupiters in systems harbouring super-Earths/sub-Neptunes with the former stating that stars with super-Earths have roughly a 3 times higher cold Jupiter fraction compared to field stars. Furthermore, they found that this cold Jupiter fraction rises to about 60\% for stars with $\rm [Fe/H]>0.1$. Given the metallicity we find for TOI-1288 of $0.07 \pm 0.09$ from the SED (median from all measurements in \tref{tab:spec_pars} is 0.15), it is perhaps not too surprising that we are seeing a cold gas giant in this system. This strong correlation between super-Earths and cold Jupiters suggests that they are not competing for the same solid material, which \citet{Zhu2018} argue disfavours theories invoking large-scale migration. 

On the other hand \rev{TOI-1288~b} is a bit larger than the planets in the aforementioned studies and might have a gaseous envelope. In line with the discussion above, hot Neptunes are in danger of losing their atmospheres, especially while their stars are young and active \citep[e.g.][]{Lopez2012}. Kozai-Lidov cycles and high-eccentricity migration can deliver Neptune-sized planets on short period orbits past this active stage ($\sim100$~Myr) for the star \citep{Dawson2018}. Interactions between TOI-1288~b and c could therefore be responsible for transporting TOI-1288~b to its current position. Subsequent tidal interactions with the star could then have dampened the eccentricity to the current value (\eb), which compared to Earth's orbit ($\sim0.016$) is still significant.

To assess whether planet c can influence the dynamics of the inner, planetary system as we see it today, we calculated the planet-star coupling parameter, $\epsilon_{\star 1}$, given in \citet{Lai2018}, which is a measure for whether an outer companion can cause the orbit of the inner planet to precess. For this we used the approximation in their Eq. (24), which is made for the case of a hot Jupiter with a gas giant companion at a separation of around 1~AU. While not exactly the case here, the approximation can still provide us with some qualitative insights.

\rev{As we} need to know the stellar rotation period, $P_{\rm rot}$, for this, we searched the \tess light curve using the autocorrelation method \citep{McQuillan2013}, however, we did not detect any signs of stellar rotation. Instead we estimated $P_{\rm rot}$ from the age, $\tau=10.05$~Gyr (\tref{tab:spec_pars}), and colour, $B-V=0.94$ (\tref{tab:star}), using the relation in \citet{Mamajek2008}, which yields a rotation period of 64~d. From this we get $\epsilon_{\star 1} \sim 0.5$ suggesting a strong coupling between TOI-1288~b and the star. However, it is not too far from the resonant regime of $\epsilon_{\star 1} \sim 1$, meaning the excitation of the spin-orbit angle, the obliquity, could be possible.

In addition to TOI-1288~c for which we have constrained the orbit and thus the mass to some extent, we also see evidence for what could be a companion on an even wider orbit. However, for the time being we can only make rather crude inferences about the characteristics of this companion as was done in \sref{sec:results}, namely \fref{fig:lower}. For instance, if this companion were on a 10~yr coplanar (with respect to TOI-1288~b) orbit it would have a mass of around 0.3~M$_{\rm J}$. To decipher the dynamic influence from this companion on the architecture would require continued RV monitoring to trace out the orbit.

\section{Conclusions}
\label{sec:conc}
Here we presented the discovery of multiple planets in the TOI-1288 system. Using photometry from \tess as well as ground-based telescopes, we have determined that the transiting planet TOI-1288~b is a super Neptune (\rpb) on a short period orbit (\Pb). TOI-1288~b thus joins the growing population of super Neptunes that despite the drought have settled in the Neptunian desert. We have characterised the planet in terms of mass through intensive RV monitoring with the HARPS-N and HIRES spectrographs, where we find a mass of \mpb.

Combining the radius and mass for TOI-1288~b, we find that the planet can be described as a rocky core with a gaseous envelope at high radiation. Similar compositions are found for the planets most identical to TOI-1288~b in terms of orbital period and radius, meaning that the internal structure and composition might be a crucial premise for survival in the desert. Atmospheric studies of occupants in and around the desert could help shed light on the processes, such as photoevaporation, shaping this region. TOI-1288~b is a well-suited candidate for such studies.

Furthermore, from our RV monitoring we also found evidence for an additional companion in the TOI-1288 system with an orbital period of \Pc. We find a lower mass of \mpc, meaning that if this companion is close to being coplanar with TOI-1288~b, it would be a Saturn-mass planet. TOI-1288~c might have been responsible for transporting TOI-1288~b from a further out orbit to its present day location, for instance, through high-eccentricity migration. Finally, we detect hints of a long-term RV trend possibly caused by another body in the TOI-1288 system.

\section*{Acknowledgements}
We thank the anonymous referee for a timely review.
This paper includes data collected by the TESS mission. Funding for the TESS mission is provided by the NASA Explorer Program. We acknowledge the use of public TESS Alert data from pipelines at the TESS Science Office and the TESS Science Operations Center.
Funding for the Stellar Astrophysics Centre is provided by The Danish National Research Foundation (Grant agreement no.: DNRF106). 
%The above might also be moved down to the other grants or the others might be moved up.
%
\rev{Based on observations (programme IDs: A40TAC\_22, A41TAC\_19, A41TAC\_49, A43TAC\_11, CAT19A\_162, CAT22A\_111, and ITP19\_1) made with the Italian Telescopio Nazionale Galileo (TNG) operated on the island of La Palma by the Fundación Galileo Galilei of the INAF (Istituto Nazionale di Astrofisica) at the Spanish Observatorio del Roque de los Muchachos of the Instituto de Astrofisica de Canarias. Based on observations made with the Nordic Optical Telescope, owned in collaboration by the University of Turku and Aarhus University, and operated jointly by Aarhus University, the University of Turku and the University of Oslo, representing Denmark, Finland and Norway, the University of Iceland and Stockholm University at the Observatorio del Roque de los Muchachos, La Palma, Spain, of the Instituto de Astrofisica de Canarias. We are extremely grateful to the NOT and TNG staff members for their unique and superb support during the observations.}
%HARPS-N observations were carried out under the programmes A41/TAC19 (P.I.~Knudstrup), A41/TACXX (P.I.~Gandolfi), A41/TACXX (P.I.~Nowak), A43/TAC11 (P.I.~Knudstrup) 
%
Resources supporting this work were provided by the NASA High-End Computing (HEC) Program through the NASA Advanced Supercomputing (NAS) Division at Ames Research Center for the production of the SPOC data products.
Some of the data presented herein were obtained at the W. M. Keck Observatory, which is operated as a scientific partnership among the California Institute of Technology, the University of California and the National Aeronautics and Space Administration. The Observatory was made possible by the generous financial support of the W. M. Keck Foundation. 
The authors wish to recognize and acknowledge the very significant cultural role and reverence that the summit of Maunakea has always had within the indigenous Hawaiian community.  We are most fortunate to have the opportunity to conduct observations from this mountain. 
This work makes use of observations from the LCOGT network. Part of the LCOGT telescope time was granted by NOIRLab through the Mid-Scale Innovations Program (MSIP). MSIP is funded by NSF. This research has made use of the Exoplanet Follow-up Observation Program (ExoFOP; DOI: 10.26134/ExoFOP5) website, which is operated by the California Institute of Technology, under contract with the National Aeronautics and Space Administration under the Exoplanet Exploration Program.
The observations in the paper made use of the NN-EXPLORE Exoplanet and Stellar Speckle Imager (NESSI). NESSI was funded by the NASA Exoplanet Exploration Program and the NASA Ames Research Center. NESSI was built at the Ames Research Center by Steve B. Howell, Nic Scott, Elliott P. Horch, and Emmett Quigley. The authors are honored to be permitted to conduct observations on Iolkam Du'ag (Kitt Peak), a mountain within the Tohono O'odham Nation with particular significance to the Tohono O'odham people. 
\rev{This work presents results from the European Space Agency (ESA) space mission \gaia. \gaia data are being processed by the \gaia Data Processing and Analysis Consortium (DPAC). Funding for the DPAC is provided by national institutions, in particular the institutions participating in the \gaia MultiLateral Agreement (MLA). The Gaia mission website is \url{https://www.cosmos.esa.int/gaia}. The \gaia archive website is \url{https://archives.esac.esa.int/gaia}.}
Some of the observations in the paper made use of the High-Resolution Imaging instrument ‘Alopeke. ‘Alopeke was funded by the NASA Exoplanet Exploration Program and built at the NASA Ames Research Center by Steve B. Howell, Nic Scott, Elliott P. Horch, and Emmett Quigley. Data were reduced using a software pipeline originally written by Elliott Horch and Mark Everett. ‘Alopeke was mounted on the Gemini North telescope of the international Gemini Observatory, a program of NSF’s OIR Lab, which is managed by the Association of Universities for Research in Astronomy (AURA) under a cooperative agreement with the National Science Foundation. on behalf of the Gemini partnership: the National Science Foundation (United States), National Research Council (Canada), Agencia Nacional de Investigación y Desarrollo (Chile), Ministerio de Ciencia, Tecnología e Innovación (Argentina), Ministério da Ciência, Tecnologia, Inovações e Comunicações (Brazil), and Korea Astronomy and Space Science Institute (Republic of Korea).
E.K. and S.A. acknowledge  the support from the Danish Council for Independent Research through a grant, No.2032-00230B.
C.M.P. and J.K. gratefully acknowledge the support of the Swedish National Space Agency (SNSA; DNR 2020-00104).
M.S.L would like to acknowledge the support from VILLUM FONDEN (research grant 42101) and The Independent Research Fund Denmark's Inge Lehmann program (grant agreement no.:  1131-00014B).
This work is partly supported by JSPS KAKENHI Grant Number JP17H04574, JP18H05439, JP20K14521, JP19K14783, JP21H00035, JST CREST Grant Number JPMJCR1761, the Astrobiology Center of National Institutes of Natural Sciences (NINS) (Grant Number AB031010).
D.H. acknowledges support from the Alfred P. Sloan Foundation and the National Aeronautics and Space Administration (80NSSC21K0652).
K.W.F.L. was supported by Deutsche Forschungsgemeinschaft grants RA714/14-1 within the DFG Schwerpunkt SPP 1992, Exploring the Diversity of Extrasolar Planets.
K.K.M. acknowledges support from the New York Community Trust Fund for Astrophysical Research.
Parts of the numerical results presented in this work were obtained at the Centre for Scientific Computing, Aarhus \url{https://phys.au.dk/forskning/faciliteter/cscaa/}.
This research made use of Astropy,\footnote{http://www.astropy.org} a community-developed core Python package for Astronomy \citep{astropy:2013, astropy:2018,astropy:2022}.
This research made use of Astroquery \citep{astroquery:2019}.
This research made use of TESScut \citep{tesscut:2019}.
This research has made use of "Aladin sky atlas" developed at CDS, Strasbourg Observatory, France \citep{Bonnarel2000,Boch2014}.
%
%%%%%%%%%%%%%%%%%%%%%%%%%%%%%%%%%%%%%%%%%%%%%%%%%%
\section*{Data Availability}

The radial velocities underlying this article are available in its online supplementary material.

%%%%%%%%%%%%%%%%%%%% REFERENCES %%%%%%%%%%%%%%%%%%

% The best way to enter references is to use BibTeX:

\bibliographystyle{mnras}
\bibliography{bibliography} % if your bibtex file is called example.bib

% Alternatively you could enter them by hand, like this:
% This method is tedious and prone to error if you have lots of references
%\begin{thebibliography}{99}
%\bibitem[\protect\citeauthoryear{Author}{2012}]{Author2012}
%Author A.~N., 2013, Journal of Improbable Astronomy, 1, 1
%\bibitem[\protect\citeauthoryear{Others}{2013}]{Others2013}
%Others S., 2012, Journal of Interesting Stuff, 17, 198
%\end{thebibliography}

%%%%%%%%%%%%%%%%%%%%%%%%%%%%%%%%%%%%%%%%%%%%%%%%%%

%%%%%%%%%%%%%%%%% APPENDICES %%%%%%%%%%%%%%%%%%%%%

\appendix

\section{Figures}

\begin{figure*}
    \centering
    \includegraphics[width=\textwidth]{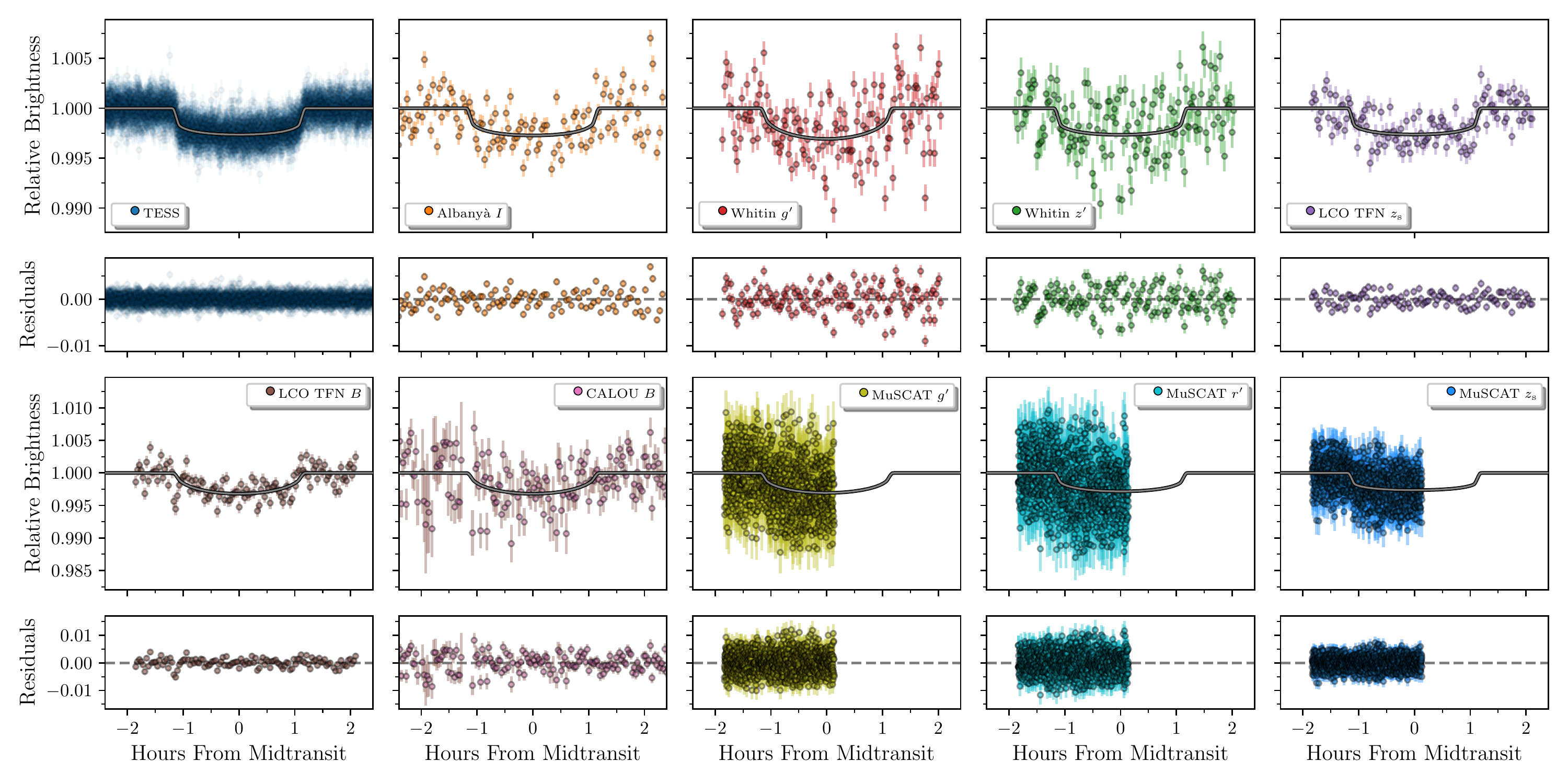}
    \caption{{\bf Light curves of TOI-1288~b.} The phase folded transits of planet b from all the different photometers. The \tess light curve (top left) is the GP detrended data from \fref{fig:tess}. The grey lines are the best-fitting models.}
    \label{fig:lcb}
\end{figure*}

%\begin{figure}
%    \centering
%    \includegraphics[width=\columnwidth]{rot_comp.pdf}
%    \caption{{\bf Sky position of blended companions.} The blue star denotes TOI-1288, while red stars are the relative positions for the companions detected in the Gemini/NIRI AO images. The orange star is the relative position of the companion detected by \gaia, which is most likely the fainter companion detected by AO.}
%    \label{fig:rot_comp}
%\end{figure}

\begin{figure}
    \centering
    \includegraphics[width=\columnwidth]{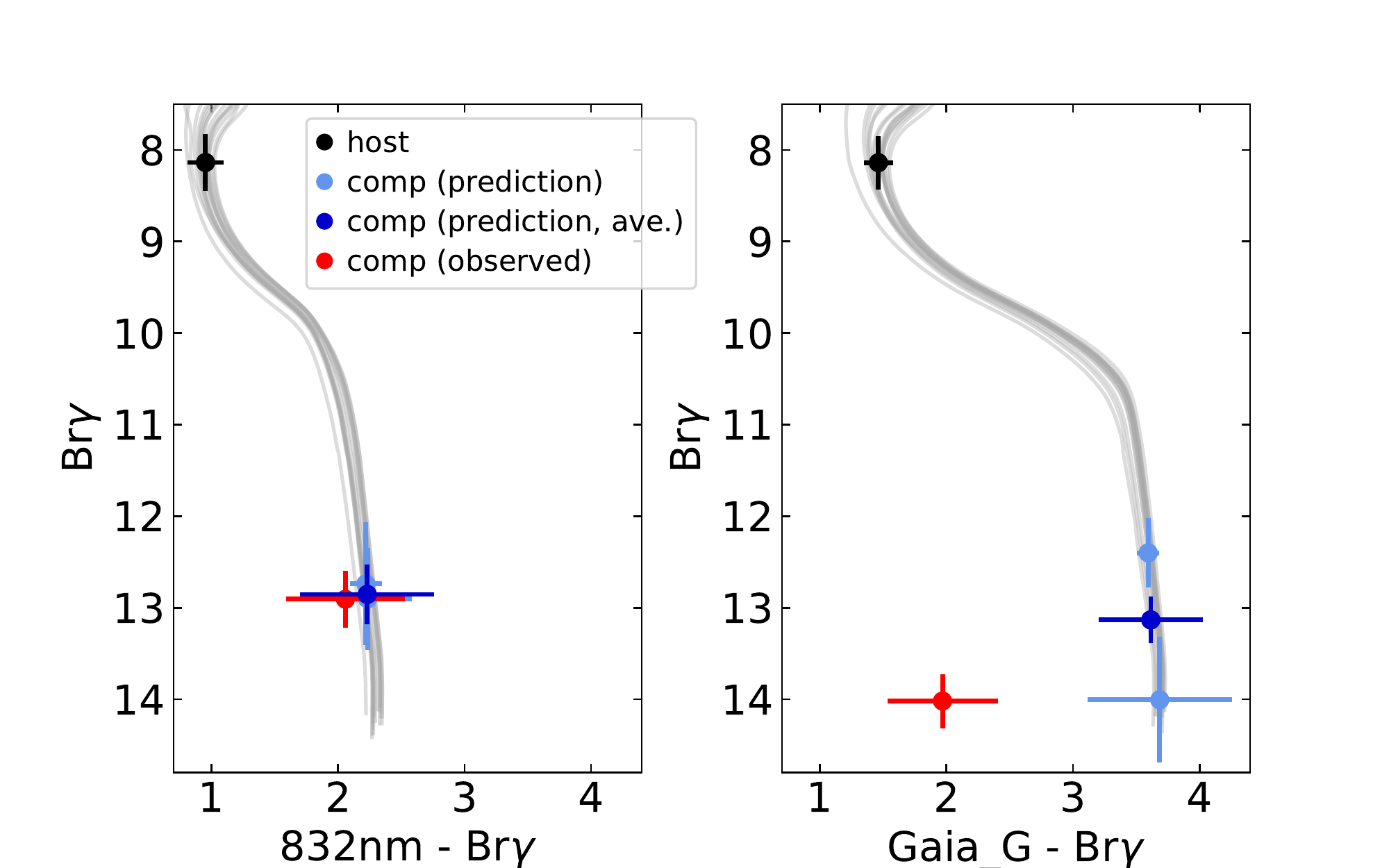}
    \caption{{\bf Colour-magnitude diagrams for TOI-1288.} Comparison of observed photometry with predicted photometry for both candidate companions. In each case, we show the CMD position of the host (black), the predicted position of a bound companion based on each measured $\Delta$mag (light blue) and the weighted average of these predictions (dark blue), and finally the observed CMD position of the companion (red). The clear disagreement for star 2 (right) indicates that this is a background object, while the relative agreement between the red and dark blue points for star 1 (left) could indicate a bound companion. As discussed in \sref{sec:bound} this is not the case.}
    \label{fig:cmd}
\end{figure}

\begin{figure}
    \centering
    \includegraphics[width=\columnwidth]{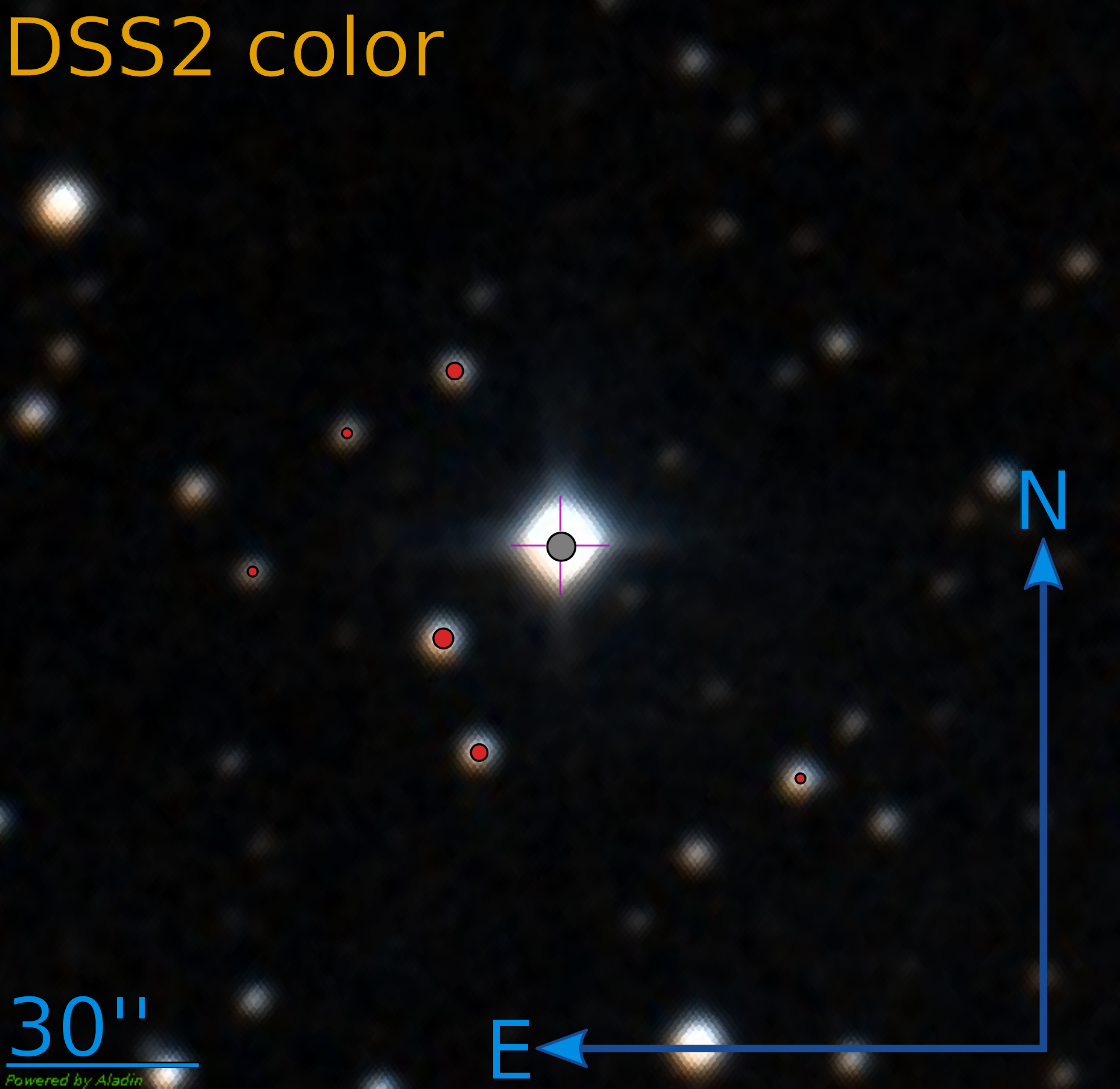}
    \caption{{\bf DSS2 image of TOI-1288.} The field around TOI-1288 as seen by the Digitized Sky Survey (DSS2). TOI-1288 is marked with the grey dot, and the red dots are the stars within the aperture in \fref{fig:tpf} using the same (relative) scaling for the marker sizes.}
    \label{fig:dss2}
\end{figure}

\begin{figure}
    \centering
    \includegraphics[width=\columnwidth]{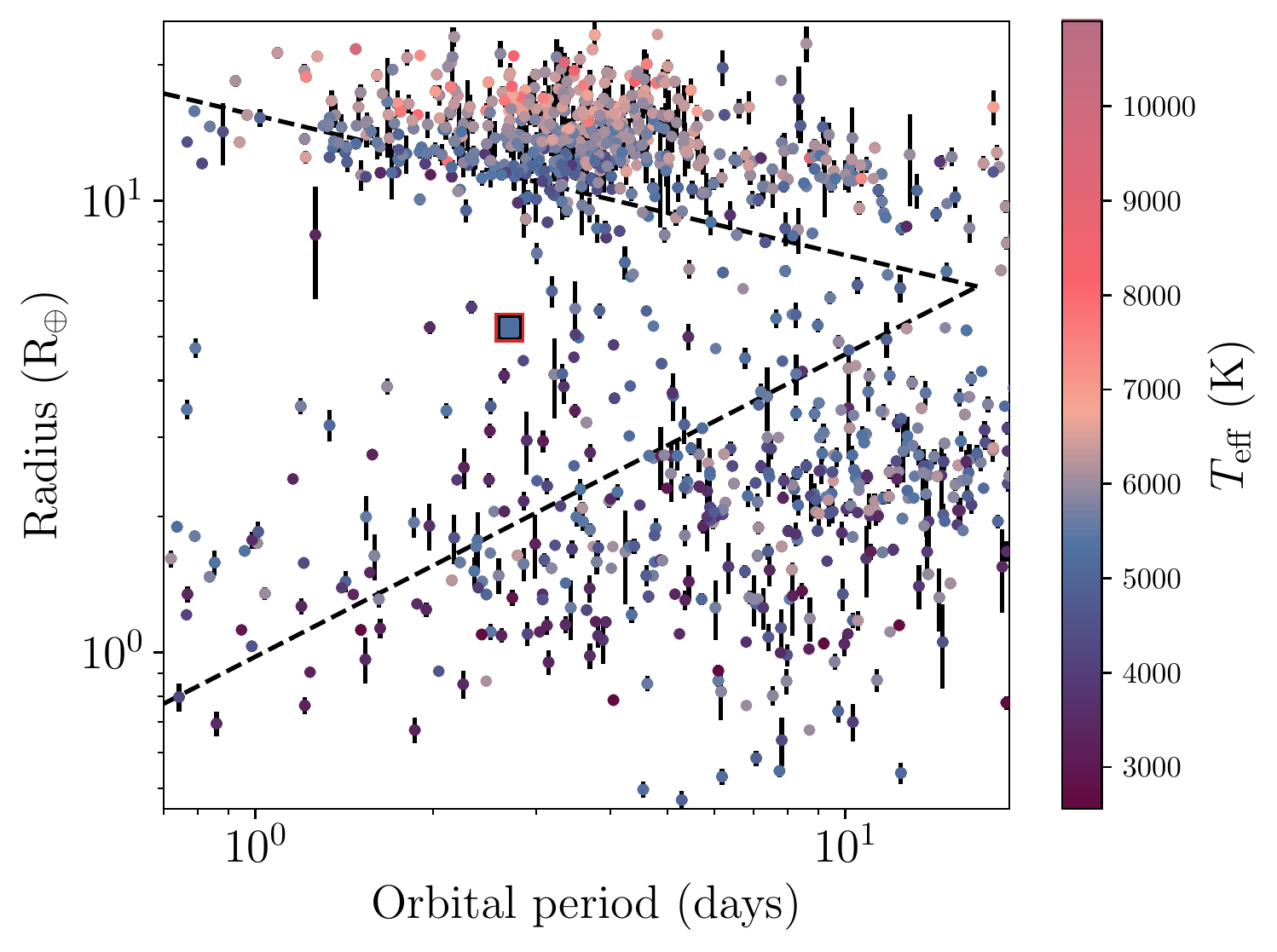}
    \caption{{\bf The hot Neptunian desert} as in \fref{fig:desert}, but with the colour coding done in the host star effective temperature.}
    \label{fig:desert_teff}
\end{figure}

\section{Tables}

\begin{table*}
\centering
\caption{{\bf Ground-based photometry.} Information on our ground-based photometric observations.}

\begin{threeparttable}

\begin{tabular}{l l c c c c c}

    \toprule
Observatory   & Location  & Aperture (m) & Photometric Aperture (arcsec)     & UTC Date      & Filter                  & Coverage  \\%&   Planet\\
    \midrule

LCOGT\tnote{1}-TFN  & Tenerife, Spain            & 1.0 & 5.8    & 2021-09-18    & $z$-short\tnote{2}              & Full \\%        & TOI-1288~b \\
LCOGT-TFN        & Tenerife, Spain       & 1.0 & 5.8   & 2021-09-18    & $B$              & Full \\%        & TOI-1288~b \\
Whitin      & Massachusetts, USA        & 0.7   & 8.0  & 2020-11-15    & $z^{\prime}$              & Full\\%         & TOI-1288~b \\
Whitin       & Massachusetts, USA       & 0.7   & 8.0  & 2020-11-15    & $g^{\prime}$              & Full\\%         & TOI-1288~b \\
Ca l'Ou\tnote{3}      & Catalonia, Spain    & 0.4 & 6.7    & 2021-05-14   & $B$              & Full   \\%      & TOI-1288~b \\
Albany$\rm \grave{a}$\tnote{4} & Catalonia, Spain   & 0.4 & 12.2  & 2019-12-02    & $I_{\rm c}$              & Full\\%         & TOI-1288~b \\
MuSCAT\tnote{5}     & Okayama, Japan         & 1.88 & 5.4    & 2019-10-31    & Sloan $g^{\prime}$              & Ingress \\%        & TOI-1288~b \\
MuSCAT     & Okayama, Japan         & 1.88  & 5.4  & 2019-10-31    & Sloan $r^{\prime}$              & Ingress  \\%       & TOI-1288~b \\
MuSCAT     & Okayama, Japan         & 1.88   & 5.4  & 2019-10-31    & $z$-short\tnote{2}              & Ingress  \\%       & TOI-1288~b \\

    \bottomrule

    \end{tabular}

\begin{tablenotes}
       \item [1] Las Cumbres Observatory Global Telescope \citep[LCOGT;][]{Brown2013}.
       \item [2] Pan-STARRS $z$-short.
       \item [3] Observatori de Ca l'Ou, Sant Martí Sesgueioles.
       \item [4] Albany$\rm \grave{a}$ Observatory.
       \item [5] Multicolor Simultaneous Camera for studying Atmospheres of Transiting exoplanets \citep[MuSCAT;][]{Narita2015}.
     \end{tablenotes}
  \end{threeparttable}
  
  \label{tab:transitfollowup}
\end{table*}

\begin{table}
    \centering
    \caption{{\bf Radial velocities.} The epochs, RVs, and errors from the HARPS-N and HIRES observations. This table is available in its entirety online.}
    \begin{tabular}{c c c c}
\toprule 
Epoch & RV & $\sigma_\mathrm{RV}$ & Instrument \\ 
BJD$_\mathrm{TDB}$ & m~s$^{-1}$ & m~s$^{-1}$ \\ 
\midrule 

2458790.340214 & -68045.93 & 1.21 & HARPS-N \\ 
2458790.359751 & -68048.67 & 1.27 & HARPS-N \\ 
2458821.330149 & -68073.51 & 1.16 & HARPS-N \\ 
\vdots & \vdots & \vdots & \vdots \\ 
2459702.697487 & -68051.86 & 0.90 & HARPS-N \\ 
2458827.739873 & 32.38 & 1.22 & HIRES \\ 
2458844.755499 & 4.83 & 1.28 & HIRES \\ 
2458852.719975 & 11.01 & 1.28 & HIRES \\ 
\vdots & \vdots & \vdots & \vdots \\ 
2459498.888082 & -14.75 & 1.51 & HIRES \\ 
\bottomrule

    \end{tabular}
    
    \label{tab:rvs}
\end{table}

\begin{table*}
    \centering
    \caption{{\bf Limb-darkening coefficients.} The limb-darkening coefficients for our MCMC using a quadratic limb-darkening law. We stepped in the sum of the coefficients with a Gaussian prior ($\mathcal{N}(\mu,\sigma)$), while keeping the difference fixed ($\mathcal{F}(c)$). The initial values were found from the tables in \citet{Claret2017} for the case of TESS and \citet{Claret2011} for the rest of the filters.}
    \begin{tabular}{c c c c}

\toprule 
Parameter & Name & Prior & Value \\ 
\midrule 
\multicolumn{4}{c}{Stepping parameters} \\ 
\midrule 
$\rm \delta M$ & Dilution  & $\mathcal{N}$(4.41,0.02) & $4.410^{+0.021}_{-0.019}$ \\ 
$(q_1 + q_2)_1$ & Sum of limb-darkening coefficients TESS & $\mathcal{N}$(0.6184,0.1) & $0.57 \pm 0.05$ \\ 
$(q_1 + q_2)_1$ & Sum of limb-darkening coefficients TESS & $\mathcal{N}$(0.6184,0.1) & $0.57 \pm 0.05$ \\ 
$(q_1 + q_2)_2$ & Sum of limb-darkening coefficients Albany$\rm \grave{a}$ $I$ & $\mathcal{N}$(0.6611,0.1) & $0.67^{+0.10}_{-0.09}$ \\ 
$(q_1 + q_2)_3$ & Sum of limb-darkening coefficients Whitin $g^{\prime}$ & $\mathcal{N}$(0.804,0.1) & $0.86^{+0.03}_{-0.07}$ \\ 
$(q_1 + q_2)_4$ & Sum of limb-darkening coefficients Whitin $z^{\prime}$ & $\mathcal{N}$(0.5544,0.1) & $0.57 \pm 0.10$ \\ 
$(q_1 + q_2)_5$ & Sum of limb-darkening coefficients LCO TFN $z$-short & $\mathcal{N}$(0.5544,0.1) & $0.53 \pm 0.09$ \\ 
$(q_1 + q_2)_6$ & Sum of limb-darkening coefficients LCO TFN $B$ & $\mathcal{N}$(0.8402,0.1) & $0.95^{+0.03}_{-0.05}$ \\ 
$(q_1 + q_2)_7$ & Sum of limb-darkening coefficients CALOU $B$ & $\mathcal{N}$(0.8402,0.1) & $0.95^{+0.03}_{-0.05}$ \\ 
$(q_1 + q_2)_8$ & Sum of limb-darkening coefficients MuSCAT $g^{\prime}$ & $\mathcal{N}$(0.804,0.1) & $0.87^{+0.03}_{-0.08}$ \\ 
$(q_1 + q_2)_9$ & Sum of limb-darkening coefficients MuSCAT $r^{\prime}$ & $\mathcal{N}$(0.712,0.1) & $0.72 \pm 0.10$ \\ 
$(q_1 + q_2)_{10}$ & Sum of limb-darkening coefficients MuSCAT $z$-short & $\mathcal{N}$(0.5544,0.1) & $0.56 \pm 0.10$ \\ 
\midrule 
\multicolumn{4}{c}{Fixed parameters} \\ 
\midrule 
$(q_1 - q_2)_1$ & Difference of limb-darkening coefficients TESS& $\mathcal{F}(0.1598)$ &  \\ 
 $(q_1 - q_2)_2$ & Difference of limb-darkening coefficients Albany$\rm \grave{a}$ $I$& $\mathcal{F}(0.2025)$ &  \\ 
 $(q_1 - q_2)_3$ & Difference of limb-darkening coefficients Whitin $g^{\prime}$& $\mathcal{F}(0.79)$ &  \\ 
 $(q_1 - q_2)_4$ & Difference of limb-darkening coefficients Whitin $z^{\prime}$& $\mathcal{F}(0.209)$ &  \\ 
 $(q_1 - q_2)_5$ & Difference of limb-darkening coefficients LCO TFN $z$-short& $\mathcal{F}(0.209)$ &  \\ 
 $(q_1 - q_2)_6$ & Difference of limb-darkening coefficients LCO TFN $B$& $\mathcal{F}(0.9002)$ &  \\ 
 $(q_1 - q_2)_7$ & Difference of limb-darkening coefficients CALOU $B$& $\mathcal{F}(0.9002)$ &  \\ 
 $(q_1 - q_2)_8$ & Difference of limb-darkening coefficients MuSCAT $g^{\prime}$& $\mathcal{F}(0.79)$ &  \\ 
 $(q_1 - q_2)_9$ & Difference of limb-darkening coefficients MuSCAT $r^{\prime}$& $\mathcal{F}(0.4344)$ &  \\ 
 $(q_1 - q_2)_{10}$ & Difference of limb-darkening coefficients MuSCAT $z$-short& $\mathcal{F}(0.209)$ &  \\ 
 \midrule 
\multicolumn{4}{c}{Derived parameters} \\ 
\midrule 
$(q_1)_1$ & Linear limb-darkening coefficient TESS &   & $0.36 \pm 0.02$ \\ 
$(q_1)_1$ & Quadratic limb-darkening coefficient TESS &   & $0.20 \pm 0.02$ \\ 
$(q_1)_2$ & Linear limb-darkening coefficient Albany$\rm \grave{a}$ $I$ &   & $0.44 \pm 0.05$ \\ 
$(q_2)_2$ & Quadratic limb-darkening coefficient Albany$\rm \grave{a}$ $I$ &   & $0.23 \pm 0.05$ \\ 
$(q_1)_3$ & Linear limb-darkening coefficient Whitin $g^{\prime}$ &   & $0.824^{+0.016}_{-0.034}$ \\ 
$(q_2)_3$ & Quadratic limb-darkening coefficient Whitin $g^{\prime}$ &   & $0.034^{+0.016}_{-0.034}$ \\ 
$(q_1)_4$ & Linear limb-darkening coefficient Whitin $z^{\prime}$ &   & $0.39 \pm 0.05$ \\ 
$(q_2)_5$ & Quadratic limb-darkening coefficient Whitin $z^{\prime}$ &   & $0.18 \pm 0.05$ \\ 
$(q_1)_5$ & Linear limb-darkening coefficient LCO TFN $z$-short &   & $0.37 \pm 0.05$ \\ 
$(q_2)_5$ & Quadratic limb-darkening coefficient LCO TFN $z$-short &   & $0.16 \pm 0.05$ \\ 
$(q_1)_6$ & Linear limb-darkening coefficient LCO TFN $B$ &   & $0.926^{+0.014}_{-0.026}$ \\ 
$(q_2)_6$ & Quadratic limb-darkening coefficient LCO TFN $B$ &   & $0.026^{+0.014}_{-0.026}$ \\ 
$(q_1)_7$ & Linear limb-darkening coefficient CALOU $B$ &   & $0.927^{+0.014}_{-0.027}$ \\ 
$(q_2)_7$ & Quadratic limb-darkening coefficient CALOU $B$ &   & $0.027^{+0.014}_{-0.027}$ \\ 
$(q_1)_8$ & Linear limb-darkening coefficient MuSCAT $g^{\prime}$ &   & $0.828^{+0.017}_{-0.038}$ \\ 
$(q_2)_8$ & Quadratic limb-darkening coefficient MuSCAT $g^{\prime}$ &   & $0.038^{+0.017}_{-0.038}$ \\ 
$(q_1)_9$ & Linear limb-darkening coefficient MuSCAT $r^{\prime}$ &   & $0.57 \pm 0.05$ \\ 
$(q_2)_9$ & Quadratic limb-darkening coefficient MuSCAT $r^{\prime}$ &   & $0.14 \pm 0.05$ \\ 
$(q_1)_{10}$ & Linear limb-darkening coefficient MuSCAT $z$-short &   & $0.38 \pm 0.05$ \\ 
$(q_2)_{10}$ & Quadratic limb-darkening coefficient MuSCAT $z$-short &   & $0.17 \pm 0.05$ \\ 
\bottomrule

    \end{tabular}
    
    \label{tab:ldcoeffs}
\end{table*}

%%%%%%%%%%%%%%%%%%%%%%%%%%%%%%%%%%%%%%%%%%%%%%%%%%

% Don't change these lines
\bsp	% typesetting comment
\label{lastpage}
\end{document}